\documentclass[fleqn,usenatbib]{mnras}

\usepackage[T1]{fontenc}

\DeclareRobustCommand{\VAN}[3]{#2}
\let\VANthebibliography\thebibliography
\def\thebibliography{\DeclareRobustCommand{\VAN}[3]{##3}\VANthebibliography}

\usepackage{amsmath,amsfonts,amssymb,amscd,wasysym,graphicx,adjustbox}

\graphicspath{{./}{figures/}}

\newcommand\xmm{\textit{XMM-Newton}}
\newcommand\nustar{\textit{NuSTAR}}
\newcommand\chandra{\textit{Chandra}}
\newcommand\wise{\textit{WISE}}
\newcommand\jwst{\textit{JWST}}

\newcommand{\ourERQfull}{SDSS J165202.60+172852.4}
\newcommand{\ourERQ}{SDSSJ1652}
\newcommand\colden{$\textrm{ cm}^{-2}$}
\newcommand\NH{$1.02$} 

\newcommand\NHwE{\NH$^{+0.76}_{-0.41}$} 
\newcommand\NHwEH{$($\NHwE$)\times10^{24}$} 
\newcommand\LX{$L_{\textrm{2-10}}$}
\newcommand\LXint{$1.4$}

\newcommand\LXintF{\LXint$^{+1}_{-1}$}
\newcommand\LXintFH{$($\LXintF$) \times10^{45}$}
\newcommand\ergs{\ $\textrm{erg s}^{-1}$}
\newcommand\flux{\ $\textrm{erg/cm}^2\textrm{/s}$}

\title[\xmm\ and \nustar\ observations of \ourERQ]{X-ray analysis of SDSS J165202.60+172852.4, an obscured quasar with outflows at peak galaxy formation epoch}

\author[Y. Ishikawa et al.]{Yuzo Ishikawa,$^{1}$\thanks{E-mail: yishika2@jhu.edu}
Andy D. Goulding,$^{2}$
Nadia L. Zakamska,$^{1}$
Fred Hamann,$^{3}$
Andrey Vayner,$^{1}$
\newauthor
Sylvain Veilleux,$^{4}$
and Dominika Wylezalek$^{5}$
\\
\\
$^{1}$Department of Physics and Astronomy, Bloomberg Center, Johns Hopkins University, Baltimore, MD 21218, USA\\
$^{2}$Department of Astrophysical Sciences, Princeton University, Princeton, NJ 08544, USA\\
$^{3}$Department of Physics and Astronomy, University of California, 900 University Avenue, Riverside, CA 92521, USA\\
$^{4}$Department of Astronomy, University of Maryland, College Park, MD 20742, USA\\
$^{5}$Zentrum f{\"u}r Astronomie der Universit{\"a}t Heidelberg Astronomisches Rechen-Institut M{\"o}nchhofstr, 12-14 69120 Heidelberg, Germany
}

\date{Accepted 2021 January 13. Received 2021 January 4; in original form 2020 August 5}

\pubyear{2021}

\begin{document}
\label{firstpage}
\pagerange{\pageref{firstpage}--\pageref{lastpage}}
\maketitle

\begin{abstract}
We report on deep \xmm\ and \nustar\ observations of the high redshift, $z=2.94$, extremely red quasar (ERQ), \ourERQfull, with known galactic ionized outflows detected via spatially-resolved [OIII] emission lines. X-ray observations allow us to directly probe the accretion disk luminosity and the geometry and scale of the circumnuclear obscuration. We fit the spectra from the \xmm/EPIC and \nustar\ detectors with a physically motivated torus model and constrain the source to exhibit a near Compton-thick column density of $N_H=$\ \NHwEH\colden, a near edge-on geometry with the line-of-sight inclination angle of $\theta_i=85^{\circ}$, and a scattering fraction of $f_{sc}\sim 3$ \%. The absorption-corrected, intrinsic 2-10 keV X-ray luminosity of \LX$=$\ \LXintFH \ergs\ reveals a powerful quasar that is not intrinsically X-ray weak, consistent with observed trends in other ERQs. We also estimate the physical properties of the obscuration, although highly uncertain: the warm ionized scattering density of $n_e \sim 7.5\times(10^2-10^3)\textrm{ cm}^{-3}$ and the obscuration mass of $M_{obsc} \sim 1.7\times(10^4-10^6) M_{\astrosun}$. As previously suggested with shallower X-ray observations, optical and infrared selection of ERQ has proved effective in finding obscured quasars with powerful outflow signatures. Our observations provide an in-depth view into the X-ray properties of ERQs and support the conclusions of severely photon-limited studies of obscured quasar populations at high redshifts.
\end{abstract}




\section{Introduction} \label{sec:intro}
Accreting supermassive black holes with $L_{bol}>10^{45} \text{ erg s}^{-1}$ -- quasars -- are of major importance to galaxy formation.
It is believed that part of the energy output is in the form of galaxy-wide outflows, impacting the host galaxy and the surrounding environment \citep{SilkRees1998AA,KormendyHo2013ARAA}. On small-scales, these outflows can be driven by the radiation pressure of quasar emission \citep{Murray1995ApJ451,Proga2000ApJ543} or by jets \citep{Fabian2012ARAA}.
The interactions between the quasar and its environments are known as quasar feedback, which is likely important in quenching star-formation, regulating central supermassive black hole growth, and limiting galaxy masses \citep{Croton2006MNRAS,Fabian2012ARAA}. 

Theoretical models suggest quasar feedback occurs at a critical phase in galaxy evolution. Quasar activity can result from accretion onto the central black hole, which can be triggered by major mergers \citep{BarnesHernquist1992,SandersMirabel1996} or more commonly through secular processes \citep{Hopkins2009ApJ694}. 
These active galaxies then enter a dusty, obscured phase enshrouded by gas and dust. Finally, the energy and momentum is released by the accreting black hole through a ``blow-out'' of galaxy-wide outflows, revealing an unobscured quasar \citep{Sanders1988ApJ,Hopkins2006ApJS}. This makes the $z\sim2-3$ epoch of high interest in galaxy formation since it marks the peak of both star formation and quasar activity \citep{Boyle&Terlevich1998MNRAS, Hopkins2006ApJS}. However, constraining the power, the mechanisms, and the impact of quasar feedback remain unresolved in galaxy formation theory \citep{Harrison2018NatAs}.

Thus, Type II \citep{Antonucci1993ARAA} quasars with powerful outflow signatures may be instrumental for probing the blow-out feedback phase. 
Of particular interest is the population of high redshift extremely red quasars (ERQs; \citealt{Ross2015MNRAS,Hamann2017MNRAS}), which are associated with extreme-velocity ionized gas outflow activity, showing strong indications of active quasar feedback in the blow-out phase \citep{Zakamska2016MNRAS4593144Z,Perrotta2019MNRAS}. Spatially resolved kinematic maps of [OIII]$\lambda$5007\AA\ emission in some of the ERQs reveal powerful galactic winds \citep{vayner20}.


To better constrain the intrinsic power and wind-driving mechanisms of the quasar's central engine, a direct measurement of the accretion luminosity of quasars with powerful outflows is needed. However, this is often  complicated by the obscuration from the optically-thick material (the torus and gas) in Type II quasars that prevent direct optical observations of the accretion disk \citep{Hickox2018ARAA}. Since powerful X-ray emission is directly associated with the corona above the quasar accretion disk, X-ray observations provide a reliable method in probing the quasar. In addition, by revealing photoelectric absorption and Compton scattering, 
X-ray measurements can provide insights into the geometry and scale of the circumnuclear obscuring material: key parameters that constrain the wind-driving mechanisms. Understanding the nature of obscuring material (scale, densities, composition, and kinematics) and the physical processes that produce the obscuration will allow us to calculate the intrinsic X-ray luminosity associated with the accretion and to probe the connection between the outflows and obscuration emerging in studies of broad absorption-line quasars \citep{Luo2013ApJ772153L}.

Uncovering and characterizing obscured quasar populations at high redshift, especially the near-Compton-thick populations, remain among the most important goals of X-ray astronomy. There is an ongoing debate in which early studies suggested that the fraction of obscured active galactic nuclei was a strongly declining function of luminosity \citep{Ueda2003ApJ}. Whether this decline is real or is due to selection effects remain controversial \citep{Lusso2013ApJ}.
\citet{LawrenceElvis2010ApJ} found no observed luminosity dependence of the obscured fraction in the radio- and mid-infrared-selected samples, and in volume-limited samples, hinting that X-ray studies may have missed Compton-thick targets. \citet{Assef2015ApJ804} suggest that at high luminosities and redshifts, there may be as many obscured and unobscured quasars. \citet{Lambrides2020ApJ} recently showed that obscured populations may be misclassified as low-luminosity objects. Therefore, it is of high interest to obtain high-quality X-ray data and to build an accurate model of the accretion properties of these unique quasars.

In this paper, we report on the joint \nustar$+$\xmm\ X-ray follow-up study of the high redshift $z=2.94$ ERQ, \ourERQfull\ (\ourERQ\ henceforth). It was identified by the Sloan Digital Sky Survey \citep{sdss2011AJ,Ross2015MNRAS} and the \textit{Wide-field Infrared Survey Explorer} (\wise; \citealt{wright2010}), and its optical and infrared properties have been well studied by our collaboration. Extreme velocities ($v>3000$ km s$^{-1}$) with strong blueshifted wings of [OIII]$\lambda\lambda$4959,5007\AA\ emission lines indicate powerful quasar-driven wind (\citealt{Alexandroff2018mnras, Perrotta2019MNRAS}); polarized UV continuum suggests dust scattering and equatorial circumnuclear outflow \citep{Alexandroff2018mnras}; and \textit{HST} imaging reveals multiple components and a kpc-scale tidal tail that suggests possible galaxy interaction \citep{Zakamska2019MNRAS}. Spatially resolved mapping of [OIII]$\lambda$5007\AA\ emission is required to prove that [OIII] winds are extended on galactic scales \citep{Greene2012ApJ746, Liu2013MNRASa,Liu2013MNRASb, Fischer2018ApJ856}. We now have direct observations that several ERQs, including SDSSJ1652, have [OIII]$\lambda$5007\AA\ winds on galactic scales from integral-field unit observations \citep{vayner20}. Finally, \ourERQ\ has been approved for the \textit{James Webb
Space Telescope} (\jwst) Early Release Science program to obtain spatially resolved spectra (ID 1335; \citealt{q3d2017jwst}), so we wish to conduct an in-depth X-ray study in preparation for \jwst.

X-ray studies of high redshift obscured quasars are often severely limited by photon statistics and rely on stacking techniques \citep{Goulding2018,Vito2018MNRAS474}. Here we present a deep X-ray study of \ourERQ\ with \xmm\ and \nustar\ conducted both to improve the knowledge of this particular object and to test the statistical techniques used in the photon-limited X-ray studies of the obscured high redshift quasar population. We measure the photon-index, $\Gamma$, of the X-ray power-law continuum ($dN/dE\propto E^{-\Gamma}$) and the absorbing column density, $N_H$, to constrain the intrinsic accretion luminosity and the geometry and scale of the obscuration. 
In Section \ref{sec:dataRedux}, we describe the observations and data reduction. In Section \ref{sec:spectAnaly}, we describe the spectral analysis. We discuss the physics of the observations in Section \ref{sec:discuss} and summarize our results in Section \ref{sec:concl}. We adopt the $h=0.7$, $\Omega_M=0.3$, and $\Omega_{\Lambda}=0.7$ cosmology. The stated uncertainties represent the 90\% confidence interval. 

\begin{table}
	\centering
	\caption{Summary of known \ourERQ\ properties. The rest-frame $L_{6\mu\textrm{m}}$ is derived from \wise\ photometry.}
	\label{tab:sdssj1652param}
	\begin{adjustbox}{max width=\columnwidth,center}
	\begin{tabular}{lccc}
	    \hline\hline
        Parameter   &   & Value       & Ref. \\ 
        \hline
        RA    &  (J2000)  & 16:52:02.60    & \citep{Gaia2018yCat}         \\
        DEC   &  (J2000)  & +17:28:52.4    &     \\
        $z$   &           & 2.94           & \citep{sdss2012DR9}         \\
        $\log L_{6\mu\textrm{m}}$      
              & (erg s$^{-1}$) & 47.19 & \citep{Goulding2018}   \\
        $\log L_{Bol}$      
              & (erg s$^{-1}$) & 47.73 & \citep{Perrotta2019MNRAS}   \\
        \hline
    \end{tabular}
    \end{adjustbox}
\end{table}

\section{Observation and data reduction} \label{sec:dataRedux}

\begin{figure*}
	 \begin{center}
	 \begin{tabular}{c}
    \includegraphics[trim={0cm 7cm 0cm 6cm}, clip,width=0.975\textwidth]{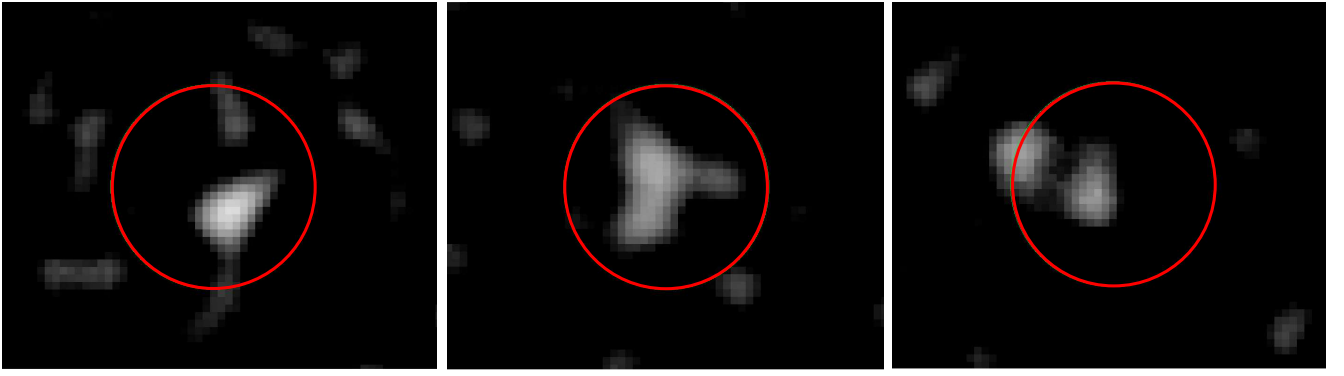}
	 \end{tabular}
	 \end{center}
	 \caption{The \xmm\ EPIC PN images, smoothed with a 5-pixel FWHM 2D Gaussian, are shown in the 0.5-2 keV, 2-5 keV, 5-10 keV bands. \ourERQ\ is centered with the circular 15'' extraction region indicated. The photometry is shown in Table \ref{tab:photometry3}.} 
	 \label{fig:j1652_img} 
\end{figure*}

\begin{table*}
    \caption{Aperture photometry results of the \xmm\ and \nustar\ observations in different energy bands. \xmm\ source counts were extracted from 15'' apertures; and the \nustar\ source counts were extracted from 30'' apertures. The EPIC detectors showed lower signal-to-noise towards the harder bands compared to that of the PN detector. Of the two \nustar\ detectors, only FPMA showed real detections, so we also show $3\sigma$ upper limits for FPMB. Only data from the EPIC and FPMA detectors were used for the spectral analysis. The observed flux has units\flux.}
	\label{tab:photometry3} 
	\begin{center}
	\begin{adjustbox}{max width=\textwidth}
	\begin{tabular}{c|ccc|ccc|ccc}
	    \hline\hline
        \xmm &   &  [0.5-2 keV] &  & &  [2-5 keV] &  &  & [5-10 keV] &  \\ 
        Detector & Net cts & Ct-rate & Flux & Net cts & Ct-rate & Flux & Net cts & Ct-rate & Flux\\ 
        \hline
        EPIC MOS-1 &
        $9.5\pm0.2$ & $1.8\times10^{-4}$ & $5\times10^{-16}$ & $1.8\pm0.3$ & $3.4\times10^{-5}$ & $4\times10^{-15}$ & $0.8\pm0.3$  & $1.5\times10^{-5}$ & $5\times10^{-15}$\\
        EPIC MOS-2 &
        $6.1\pm0.8$ & $1.3\times10^{-4}$ & - & $7.3\pm0.5$ & $1.6\times10^{-4}$ & - & $3.3\pm0.6$  & $7.1\times10^{-5}$ & -\\ 
        EPIC PN &
        $15.0\pm0.4$ & $3.1\times10^{-4}$ & - & $33.0\pm0.5$ & $6.8\times10^{-4}$ & - & $39.4\pm0.6$ & $8.1\times10^{-4}$ & -\\ 
	    \hline
	    \hline
	    \nustar  &   &  [6-10 keV] & &  &  [10-40 keV] &   &  &[40-79 keV]  & \\ 
        Detector & Net cts & Ct-rate & Flux & Net cts & Ct-rate & Flux & Net cts & Ct-rate & Flux\\ 
        \hline
        FPMA & $22\pm0.7 $ & $9.9\times10^{-5}$ & $4\times10^{-15}$ & $25\pm1$ & $11\times10^{-5}$ & $1\times10^{-14}$ & $13\pm1$ & $5.9\times10^{-5}$ & $7\times10^{-15}$\\
        FPMB & $<2$  & $<9.2\times10^{-6}$ & - & $<4$  & $<1.9\times10^{-6}$ & - & $<3$  & $<1.4\times10^{-6}$ & -\\ 
        \hline
    \end{tabular}
    \end{adjustbox}
	\end{center}
\end{table*}

\begin{table*}
    \caption{The \texttt{XSPEC} commands for each model considered. A constant Galactic absorption (\texttt{phabs}) is assumed for both models. The \texttt{MYTorus} model describes a reprocessed spectrum (\texttt{MYTZ}, \texttt{MYTS}, and \texttt{MYTL}) and an optically thin scattered continuum, assuming a terminal energy of 500 keV. Normalization for the Compton scattered emission (\texttt{const$_2$}) and fluorescent line emission (\texttt{const$_3$}) are tied together for a self-consistent model. The scattering fraction $f_{sc}$ is given by \texttt{const$_4$}, which is linked to the power-law model of the source \texttt{zpowerlw}.}
    \begin{tabular}{cl}
        \hline\hline
        Model               &\texttt{XSPEC} command \\
        \hline
        Absorbed power-law  &\texttt{phabs*zphabs*zpowerlw} \\
        \texttt{MYTorus}&\texttt{const$_1$*phabs(zpowerlw*MYTZ+const$_2$*MYTS}
                        \texttt{+const$_3$*gsmooth*MYTL+const$_4$*zpowerlw)} \\
        \hline
    \end{tabular}
    \label{tab:modelList} 
\end{table*}


\begin{figure*}
	 \begin{center}
	 \begin{tabular}{c}
    \noindent\includegraphics[trim={0.4cm 0.5cm 0.2cm 0.2cm}, clip, width=0.977\textwidth] {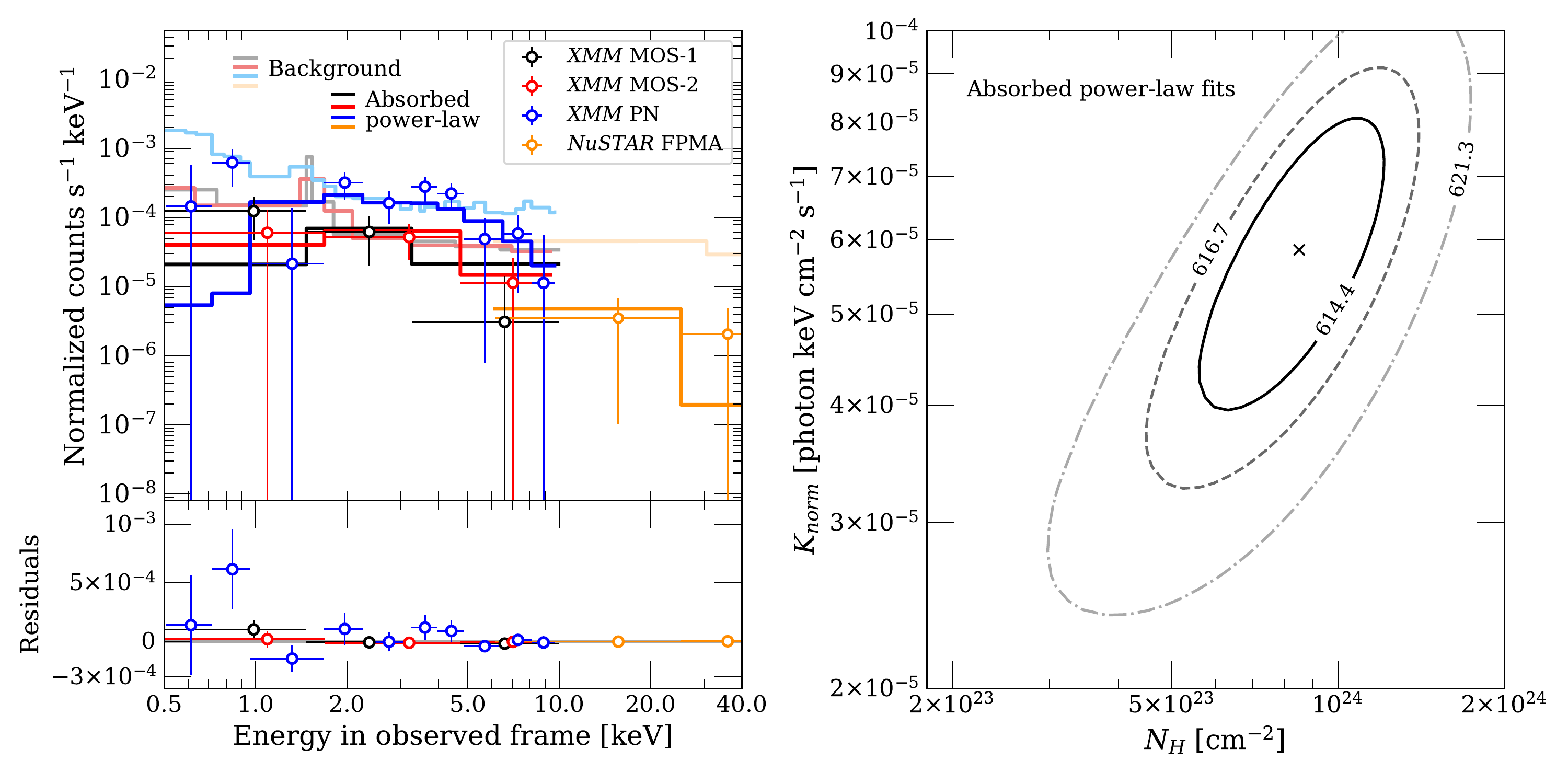}
	 \end{tabular}
	 \end{center}
	 \caption{\textbf{(Left top)} The \xmm\ and \nustar\ spectra and the folded absorbed power-law model fits are shown. The spectra have been rebinned for plotting purposes. The binned data points are indicated with circles, and the models are shown as solid lines: MOS-1 (black), MOS-2 (red), PN (blue), and FPMA (orange). The extracted background spectrum for each detectors are also shown as faded lines in the same color scheme.  \textbf{(Left bottom)} The data-model residuals show increased deviations in the $< 1.5$ keV soft energy region, which indicates the need for a more complex model. While it is possible that the soft energy features may be background driven, it is unlikely that features around 3-5 keV are affected by the background. \textbf{(Right)} The confidence contours for the best-fitting $K_{norm}$ and $N_H$ for the absorbed power-law model. The contour lines represent the $68 \%$ (solid), $90\%$ (dashed), and $99\%$ (dashed-dot) $C_{stat}$ confidence levels for 693 degrees of freedom. We see that the model indicates a near-Compton thick absortion of $N_H\sim10^{24}\textrm{ cm}^{-2}$.}
	 \label{fig:simple_model_wContour} 
\end{figure*}

\subsection{\xmm}
\ourERQ\ was observed with \xmm\ \citep{Jansen2001AA} on 2019 March 5-6 for an exposure of 130 ks (ObsID: 0830520101). Exposures were taken with the European Photon Imaging Camera (EPIC), which carries a set of three X-ray CCD cameras: MOS-1, MOS-2, and PN \citep{struder2001AA,Turner2001AA}. The EPIC CCDs are sensitive from 0.15-15 keV with moderate spectral energy resolution ($E/\Delta E \sim$ 20-50) and point-spread function with 6'' full width at half maximum (FWHM). 

Each EPIC images was reduced with the \textit{XMM-Newton Science Analysis Software} (XMMSAS; \citealt{Gabriel2004xmmsas}) v18.0.0 and the Current Calibration Files (CCF) v3.12. The calibrated events were produced using the \texttt{EPPROC} task. A rate-filter was applied to remove minor background flares, such that the effective exposure time was cut to approximately 50 ks. 

We extracted the source spectra from 15'' radius apertures. Background spectra were extracted from the same chip as the source; however the aperture was treated differently depending on the source's location on the chip. For the EPIC-PN, we took a circular 110'' radius aperture near the source aperture. For the EPIC-MOS, we took a 60'' to 200'' annular aperture centered on the source. Since the background regions also overlapped with neighboring point-source detections, we excluded 15'' radii apertures for each contaminant. 
Images of the source with the 15'' extraction apertures for the PN detector at three energy bands are shown in Figure \ref{fig:j1652_img}. 
The redistribution matrix file (RMF) and auxiliary response file (AMF) were generated with the \texttt{RMFGEN} and \texttt{ARFGEN} respectively. The final background-subtracted spectra were grouped at 1 photon per bin using \texttt{grppha}. Photometry results are summarized in Table \ref{tab:photometry3}.

\subsection{\nustar}
The \nustar\ \citep{Harrison2013ApJ770} observatory is the first focusing satellite with sensitivity over a broad high-energy 3-79 keV energy band. It consists of two focal-plane modules (FPMA and FPMB) with a 12'$\times$12' field of view. Since we expected low-photon detections due the predicted Compton-thick nature of the \ourERQ, a much longer 218 ks observation was performed with \nustar\ on 2019 February 23 (obsID: 60401028002). The \nustar\ data were reduced with the \textit{NuSTAR Data Analysis Software} (NuSTARDAS) v20160502 and the \textit{NuSTAR Calibration Database} (CALDB), using the \texttt{nupipeline} tasks. The spectra and light-curves were extracted with the \texttt{nuproducts} task.

Photometry results in Table \ref{tab:photometry3} were extracted from a 30'' radius circular region centered on the target coordinates from both detectors. Despite the long exposure, the FPMB detector did not detect any significant signal from the source. Thus we only use data from the FPMA detector for spectral analysis. We calculate the 3$\sigma$ upper limit for FPMB in three bands: 6-10 keV, 10-40 keV, and 40-79 keV. The \nustar\ photometry results are shown in Table \ref{tab:photometry3}. The non-detection can be attributed to a few reasons: high attenuation of X-ray signals due to the obscuring material, limited exposure time, and stray light, which complicated background subtraction. 

\section{X-ray spectral analysis}\label{sec:spectAnaly}

\begin{table*}
    \caption{Best-fit parameters for the unabsorbed power-law (PL), absorbed power-law model (Abs. PL), and the \texttt{MYTorus} models at varying $\theta_i$. The power-law index is fixed at $\Gamma=1.9$ for all models with absorption. For the detailed error bounds, refer to the confidence contours in Figures \ref{fig:simple_model_wContour} and \ref{fig:MYTorus_contour} for the absorbed power-law and the \texttt{MYTorus} models, respectively. $L_{\textrm{2-10},obs}$ indicate observed luminosity, and $L_{\textrm{2-10},int}$ indicate absorption-corrected luminosity, both in rest-frame. All parameters have been redshift corrected. Stated uncertainties represent the 90\% confidence interval.}
	\label{tab:bestFit} 
	\begin{tabular}{lccccccc}
	    \hline\hline
        Parameter & & PL & Abs. PL & \texttt{MYTorus}      & & &  \\ 
        & & & & $\theta_i=45^{\circ}$ & $=60^{\circ}$ & $=75^{\circ}$ & $=85^{\circ}$ \\
        \hline
        $\Gamma$ & & $0.8\pm0.3$ & (1.9) & (1.9) & (1.9) & (1.9) & (1.9) \\ 
        
        $K_{norm}$ & ($10^{-5}$ phot/keV/cm$^2$/s)  &
        $0.1^{+0.2}_{-0.1}$ &
        $6\pm2$ &
        $0.3^{+0.3}_{-0.9}$ & $0.9^{+0.3}_{-0.9}$ & 
        $8^{+7}_{-4}$ & $9^{+7}_{-4}$ \\ 
        
        
        $N_H$ & ($10^{24}$\colden)  & 
        - &
        $0.9^{+0.4}_{-0.3}$ &
        $2\pm52$ & $2\pm27$ &  
        $1.2^{+1}_{-0.5}$ & \NHwE \\ 
        
        $f_{sc}$ & $(10^{-2})$ & - &
        - &
        $3\pm20$ & $1\pm4$ &
        $3.5\pm2.8$ & $3.6\pm2.7$ \\
        
        C-stat & (dof) & 617 (665) & 633 (693) & 656 (692) & 655 (692) & 610 (664) & 610 (667)  \\
        \hline
        $L_{\textrm{2-10},obs}$ & ($10^{44}$ erg s$^{-1}$)  &
        - &
        $0.7^{+0.2}_{-0.2}$ &
        $1^{+2}_{-0.9}$  & $1^{+4}_{-0.9}$ &
        $0.9^{+0.3}_{-0.2}$ & $0.9^{+0.2}_{-0.3}$   \\ 
        
        $L_{\textrm{2-10},int}$ & ($10^{45}$ erg s$^{-1}$)  &
        $0.11^{+0.01}_{-0.05}$ &
        $0.6^{+0.4}_{-0.2}$ &
        $2.5^{+2}_{-1}$ & $0.9\pm0.9$ &
        $1.2\pm1$ & \LXintF   \\ 
	    \hline
    \end{tabular}
\end{table*}

We perform spectral analysis on \ourERQ\ with \texttt{XSPEC} v12.10 \citep{Arnaud1996ASPC} for photon-limited Cash statistics \citep{Cash1979ApJ}. We adopt the Galactic absorption column density of $N_{H,Gal}=5.03\times10^{20}\textrm{ cm}^{-2}$ \citep{HI4PI2016AA}, which is modeled by the \texttt{phabs} \texttt{XSPEC} model. We perform simultaneous spectral analysis on the \xmm\ EPIC (MOS and PN) between 0.5-10 keV and the \nustar\ FPMA data between 6-79 keV, excluding the FPMB data due to low signal. The objectives are to constrain the power-law photon index $\Gamma$ associated with the intrinsic luminosity and the column density $N_H$ to understand the geometry and scale of the obscuration. All fits are performed in the rest-frame assuming $z=2.94$. 

\subsection{Absorbed power-law model}
We first fit the \ourERQ\ spectra with a power-law. 
The initial fit produces a small photon-index $\Gamma\sim0.8$. Since the typical value of the power-law slopes measured in active galactic nuclei is $\Gamma\approx2$ \citep{NandraPounds1994MNRAS, Reeves2000MNRAS, Page2005MNRAS,Piconcelli2005AA}, the flat spectral slope of $\Gamma\sim0.8$ is unlikely to reflect the intrinsic properties of the source, but rather indicates the presence of significant absorption \citep{Alexander2001AJ}. The resulting observed spectrum is harder than the intrinsic spectrum of an unobscured quasar \citep{NandraPounds1994MNRAS,Goulding2018}. 

Next, we model the spectra using a power-law with photoelectric absorption in the rest-frame. We refer to this as the absorbed power-law model, and the corresponding \texttt{XSPEC} model is listed in Table \ref{tab:modelList}. When all model parameters (column density $N_H$, normalization $K_{norm}$, and the photon index $\Gamma$) are set as free parameters, the best-fit model also produce a low power-law index $\Gamma\sim0.8$, which is likely due to the high soft-energy excess, hinting complex attenuation mechanism. Instead, we adopt the approach taken by \citet{Goulding2018}, among others, and force a more physically motivated model. We fix the photon index at a typical slope of $\Gamma=1.9$ and freely fit for the column density $N_H$ and energy normalization $K_{norm}$. The observed spectra with the best-fit model are shown in Figure \ref{fig:simple_model_wContour}. The best-fit parameters are $N_H=(9^{+4}_{-3})\times10^{23}\textrm{ cm}^{-2}$ and $K_{norm}=(6\pm2)\times10^{-5} \textrm{ phot keV/cm}^2\textrm{/s}$ at 1 keV with an improved $C_{stat}=633$ for 693 degree of freedom (dof). The confidence ranges of the calculated $N_H$ and $K_{norm}$ are produced with the \texttt{XSPEC} \texttt{steppar} routine. The confidence contours of $N_H$ and $K_{norm}$ fits is shown in Figure \ref{fig:simple_model_wContour}. The power-law model with $\Gamma=1.9$ yields the same results for the intrinsic luminosity whether it is applied to just the \nustar\ data or to the combined \xmm$+$\nustar\ dataset. 
Spectral fit results are summarized in Table \ref{tab:bestFit}.

The best-fitting model spectrum in Figure \ref{fig:simple_model_wContour} displays a noticeable inversion around the observed-frame 1.5 keV. Examining the data-model residual, we see that this absorbed power-law model does not capture this observed-frame soft energy excess in the 0.5-1 keV range. Although the Fe K $\alpha$ emission line (1.6 keV in the observed-frame) is present \citep{Piconcelli2015AA}, due to the low counts observed, we cannot find an appropriate binning that comfortably shows both the line and the continuum. These observed spectral features combined with the initial column density estimate approaching $N_H\sim 10^{24}\textrm{ cm}^{-2}$ suggest significant Compton-thick obscuration with a more complex absorption structure.

\begin{figure*}
	 \begin{center}
	 \begin{tabular}{c}
	\includegraphics[trim={0.4cm 0.5cm 0.35cm 0.25cm},clip, width=0.9\textwidth] {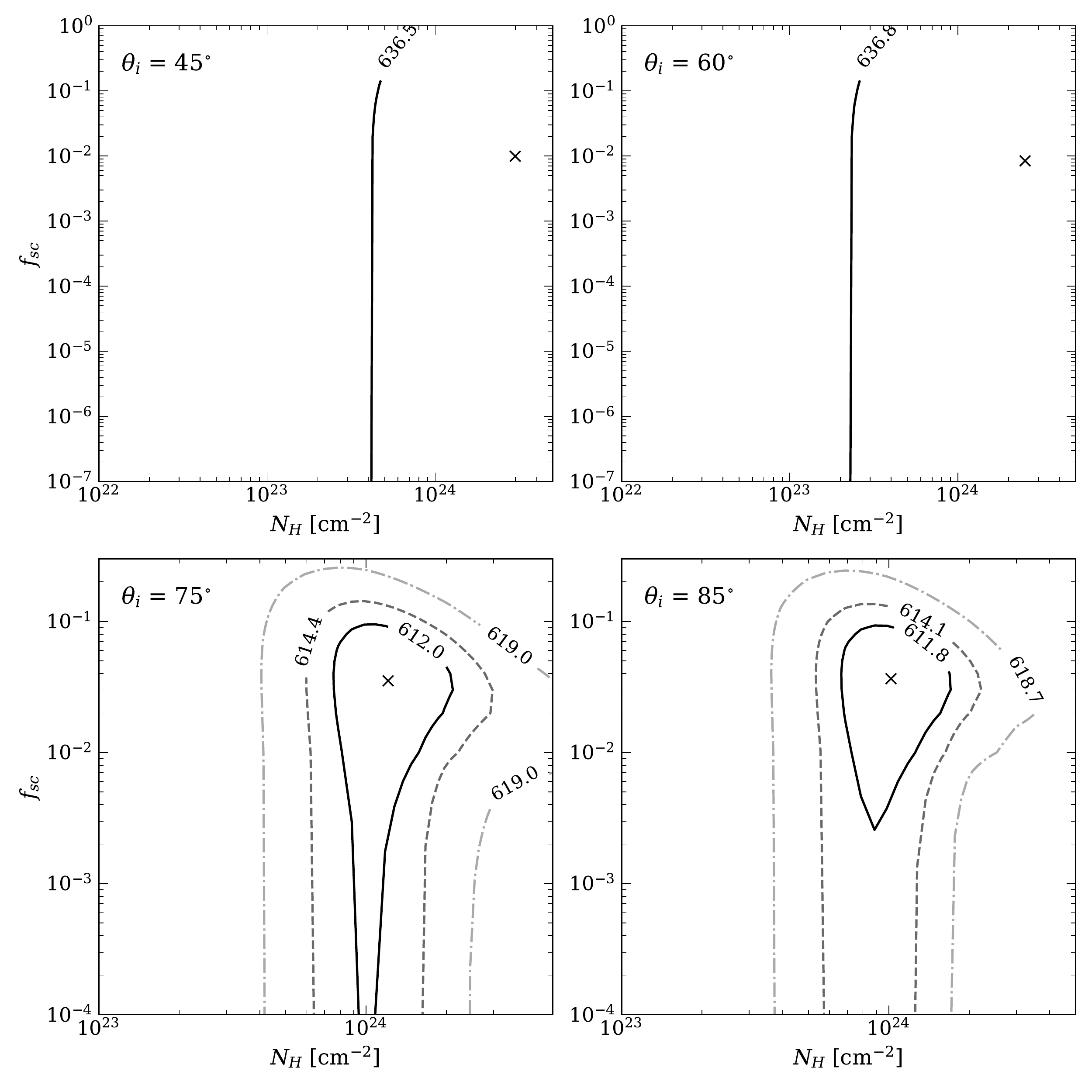}
	 \end{tabular}
	 \end{center}
	 \caption{The confidence contours for the best-fitting \texttt{MYTorus} parameters $N_H$ and $f_{sc}$ for $\theta_i=45^{\circ}$ (top-left), $60^{\circ}$ (top-right), $\theta_i=75^{\circ}$ (bottom-left), and $85^{\circ}$ (bottom-right) at $68 \%$ (solid), $90\%$ (dashed), and $99\%$ (dashed-dot) $C_{stat}$ confidence levels. It was not possible to constrain the confidence contours better than $68 \%$ for $\theta_i=45^{\circ}$ (left) and $60^{\circ}$. Comparing the contours, we find that larger inclinations are preferred; and the $\theta_i=85^{\circ}$ model appear to better constrain $N_H$ and $f_{sc}$ over $\theta_i=75^{\circ}$ .}
	 \label{fig:MYTorus_contour} 
\end{figure*}


\begin{figure*}
    \begin{center}
    \begin{tabular}{c}
        \includegraphics[width=0.977\textwidth]{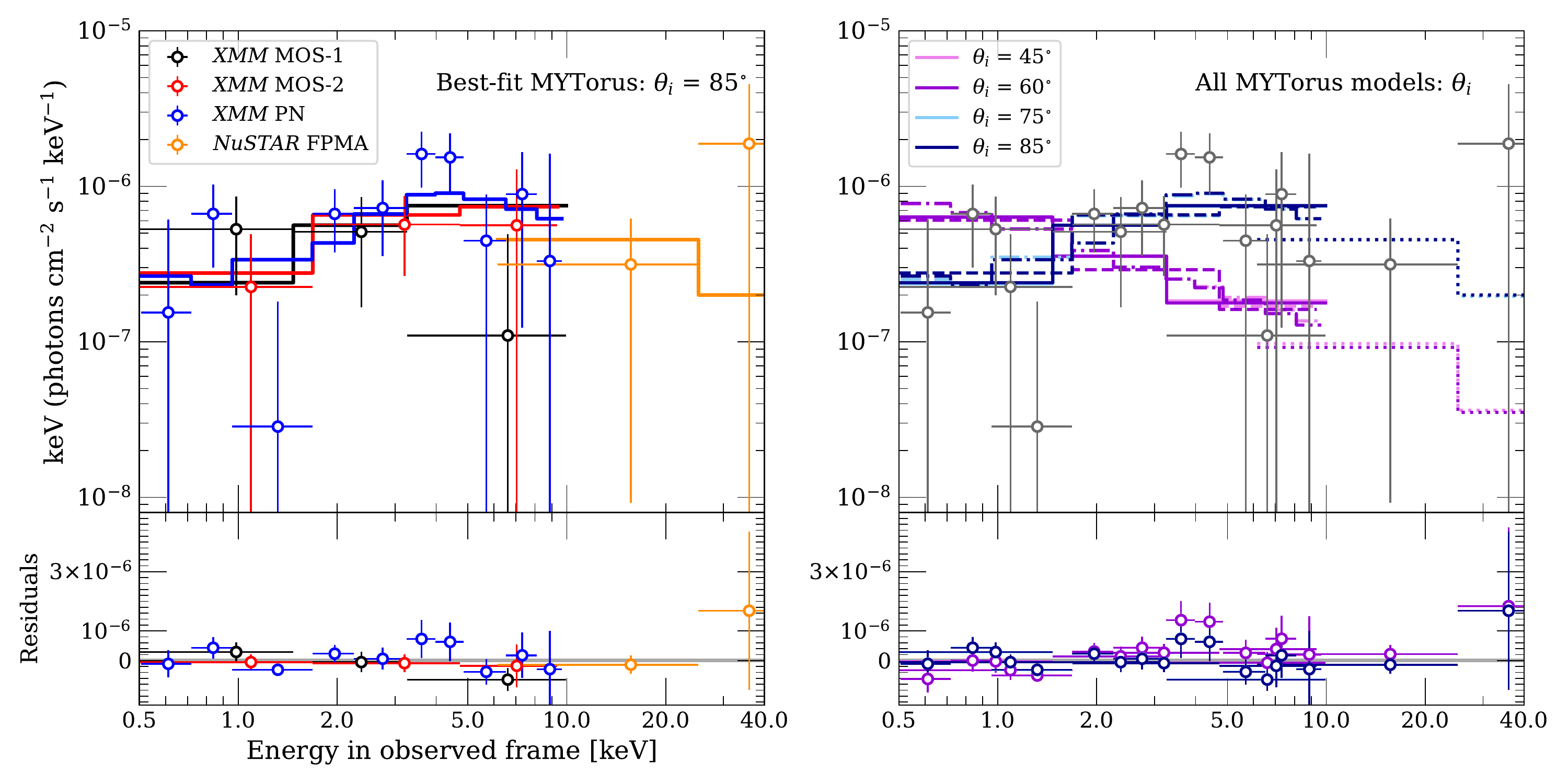}\\
        \includegraphics[width=0.97\textwidth]{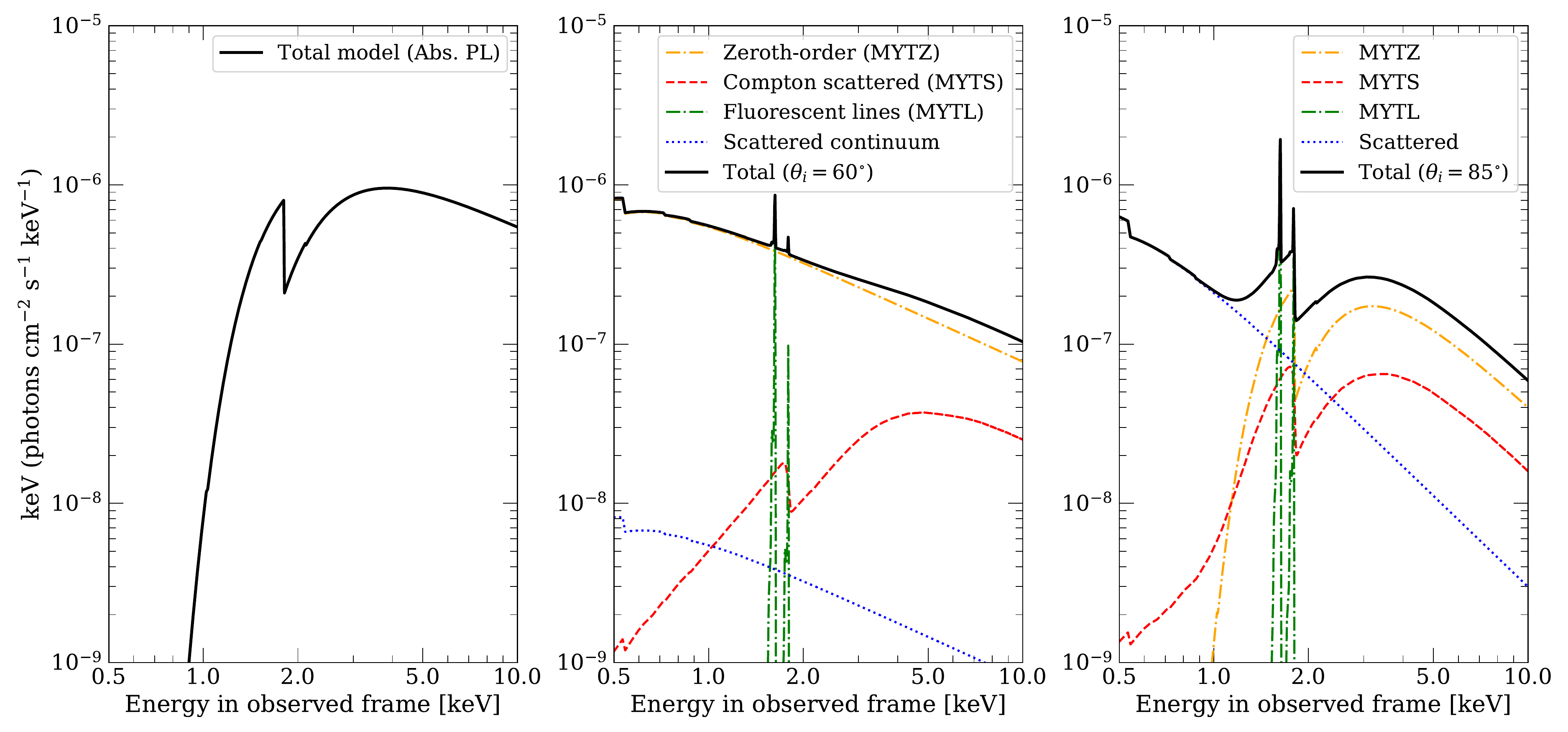}
    \end{tabular}
    \caption{\textbf{(Top left)} The observed-frame, unfolded energy spectrum with the best-fit $\theta_i=85^{\circ}$ \texttt{MYTorus} model. The spectra have been re-binned for plotting purpose. The binned data points are shown in circles, and the black, red, blue, and orange solid lines represent the best-fit \texttt{MYTorus} model at $\theta_i=85^{\circ}$.  \textbf{(Top right)} All of the \texttt{MYTorus} models for the various $\theta_i$ considered are plotted against the binned data points (in gray). The data-model residuals in units of keV are also shown and show that models at larger $\theta_i$ (light/dark blue) capture both the soft and hard energy data better than at smaller $\theta_i$ (light/dark purple). \textbf{(Bottom panel)} Comparison of the observed frame absorbed power-law (left) and \texttt{MYTorus} models, $\theta_i=60^{\circ}$ (center) and $\theta_i=85^{\circ}$ (right), with the power-law parameter fixed at $\Gamma=1.9$. The solid black line indicates the total model spectrum, while the colored dashed/dotted lines indicate the individual model components. We note that a pure absorbed power-law does not capture the soft-energy components, while the less inclined \texttt{MYTorus} $\theta_i=60^{\circ}$ spectrum predicts a smoother profile without a prominent hard Compton scattered continuum past the Fe K $\alpha$ fluorescent line, suggesting a highly inclined geometry such as the $\theta_i=85^{\circ}$ model. Refer to Table \ref{tab:bestFit} for the best-fit parameters of each model.}
    \label{fig:allMYTorusSpectra}
    \end{center}
\end{figure*} 

\subsection{\texttt{MYTorus} model}
We adopt the physically motivated \texttt{MYTorus} spectral model \citep{MY09MNRAS,Yaqoob2012MNRAS}, which is a Monte Carlo model based on a toroidal circumnuclear reprocessor that also accounts for geometry. The increased soft X-ray component seen in Figure \ref{fig:simple_model_wContour} suggests a Compton scattering continuum, which \texttt{MYTorus} can also model self-consistently. We fit a model that uses pre-calculated \texttt{MYTorus} tables that include: (a) the zeroth-order reprocessed power-law continuum - \texttt{MYTZ}, (b) the Compton-scattered continuum - \texttt{MYTS}, (c) the scattered fluorescent line emission due to Fe K at $\sim6.4\textrm{ keV}$ rest-frame - \texttt{MYTL}, and (d) the optically-thin scattered power-law continuum, separate from the \texttt{MYTorus} tables, to account for electron scattering in a warm/hot ionized region surrounding the central engine. This added optically-thin scattering region is larger than the obscuring structure. For simplicity, we refer to this model set as the \texttt{MYTorus} model throughout the paper. This spectral shape is motivated by the stacked analysis in \citet{Goulding2018}. The \texttt{XSPEC} implementation of this model is shown in Table \ref{tab:modelList}. All models considered assume that the intrinsic power-law continuum terminates at 500 keV, referred to as the terminal energy.

Similar to the absorbed power-law model treatment, we fix $\Gamma=1.9$ and tie it to each \texttt{MYTorus} model components. We also fix the \texttt{MYTorus} normalization constants at unity and link each of their column densities $N_H$. Previous UV spectropolarimetry analysis suggested a broad-line region scattering, in which the likely line-of-sight inclination angle $\theta_i$ to the central engine is near edge-on; however, $\theta_i$ was not constrained at the time \citep{Alexandroff2018mnras}. Rather than treating $\theta_i$ as another free parameter, we fix at a particular value, and freely fit for $K_{norm}$, $N_H$, and the scattering fraction $f_{sc}$ corresponding to the optically-thin scattered continuum. We repeat this fitting process over four angles: $\theta_i=45^{\circ}$, $60^{\circ}$, $75^{\circ}$, and $85^{\circ}$ (edge-on is defined as $90^{\circ}$). In each iteration of the fit, we initially fix $f_{sc}=1\%$ to localize $N_H$, then free the $f_{sc}$ parameter for simultaneous $K_{norm}$, $N_H$, and $f_{sc}$ fitting.

The best-fit parameters for the four \texttt{MYTorus} models are listed in Table \ref{tab:bestFit}, and the corresponding confidence contour maps of the $N_H$ and $f_{sc}$ fits are shown in Figure \ref{fig:MYTorus_contour}. We find a strong dependence on $\theta_i$, in which greater inclination angles are preferred. In fact, the uncertainties for the $\theta_i=45^{\circ}$ and $60^{\circ}$ could not be constrained, reaching extreme estimates for $N_H$ and $f_{sc}$ as indicated by the $68 \%$ contour. From the fit values and the contour maps, we conclude that the $\theta_i=75^{\circ}$ and $85^{\circ}$ geometries fit the data better. Since the estimated values for $N_H$ and $f_{sc}$ at these two angles are nearly identical with overlapping error bounds, it is difficult to conclusively pick one model over the other. Neither were we able to constrain a lower limit for $f_{sc}$ to high confidence. However, the confidence contour maps in Figure \ref{fig:MYTorus_contour} suggest that the model parameters are better constrained at  $\theta_i=85^{\circ}$ with $N_H=$\ \NHwEH\colden\ and $f_{sc}= (3 \pm 2)$ \% at $90 \%$ confidence with $C_{stat}=610$\ for 667 degrees of freedom. 

The unfolded energy spectrum with the best-fit $\theta_i=85^{\circ}$ \texttt{MYTorus} model is shown in Figure \ref{fig:allMYTorusSpectra}. We compare the different best-fitting unfolded \texttt{MYTorus} $\theta_i$ models with the observed data. We also plot the models and their respective components in Figure \ref{fig:allMYTorusSpectra}. Despite the differences in fits, all of the \texttt{MYTorus} model fit results suggest that \ourERQ\ is a highly obscured, Compton-thick quasar. The calculated model (redshift and absorption-corrected) rest-frame luminosity for the $\theta_i=85^{\circ}$\ model is \LX$=$\ \LXintFH \ergs.  


\subsection{Comparing the models}
The C-stat values of the best fitting the absorbed power-law and \texttt{MYTorus} models both indicate similar goodness-of-fits. Here we focus on the data-model residuals in Figures \ref{fig:simple_model_wContour} and \ref{fig:allMYTorusSpectra}. In general, the \texttt{MYTorus} model-fits capture the spectral features across all energies better than absorbed power-law model (Figure \ref{fig:simple_model_wContour} left vs. Figure \ref{fig:allMYTorusSpectra} top left). We argue that the \texttt{MYTorus} model fit is mostly driven by the spectral shape in $E>2$ keV observed frame. In fact, according to just the C-stat values, the $\theta_i=45^{\circ}$ and $60^{\circ}$ \texttt{MYTorus} models would be preferred over the $\theta_i=75^{\circ}$ and $85^{\circ}$ models, which is not supported by the parameter confidence contours. An obscuring structure with a more inclined geometry is preferred to explain the medium/hard energies.

However, there are some uncertainties within the \texttt{MYTorus} model fits that do not fully model the spectral features at $<1.5\textrm{ keV}$ and $\textrm{3-5 keV}$, which are more pronounced in the \xmm/PN data (blue data points). The background spectrum in Figure \ref{fig:simple_model_wContour} do not show any significant features at 3-5 keV, suggesting that the unmodeled features are not likely to be background driven. Unmodeled components in 3-5 keV may suggest that we are not fully modeling the peak of the X-ray emission, implying underestimated $N_H$, \LX, and $f_{sc}$. 
 
To understand the soft energy fits, we also considered \texttt{MYTorus} models without the warm electron scattering, which produce similar best-fit $\theta_i$, $N_H$, $K_{norm}$, and C-stat values. With this modification, the best-fit spectrum  displays deviations in the soft-energies similar to that of the absorbed power-law model, suggesting the need for a soft-energy scattering component. Therefore, we argue that the inclined \texttt{MYTorus} geometry with the scattering region is the preferred model. A possible explanation for the systematic model residuals in our best-fit \texttt{MYTorus} model is a presence of a more complex thermal structure, not captured by \texttt{MYTorus} or the electron scattering components. However, adding a thermal component did not change the parameters nor improve the fit quality, so we cannot preferentially constrain the model.  Another possibility is a presence of a more complex obscuring torus that can be described with variable opening angle model (\texttt{borus}; \citealt{Balokovic2018}) as explored in \cite{LaMassa2019ApJ}, but this would require higher signal data. Finally, the remaining uncertainties may reflect the poor data quality from severely photon limited studies. 


\section{Discussion}\label{sec:discuss}

\subsection{Power of central engine}
There is an ongoing debate about the relationship between outflows and accretion in luminous quasars: whether near-Eddington accretion is associated with disrupting the X-ray emitting corona or whether the corona must be compact and shielded to enable radiative driving \citep{Luo2013ApJ772153L}. The X-ray luminosity is one of the key determinants of the wind launching mechanism.

\begin{figure}
	 \begin{center}
    	 \begin{tabular}{c}
    	\includegraphics[trim={0.45cm 0.5cm 0cm 0cm}, clip,width=0.952\columnwidth]{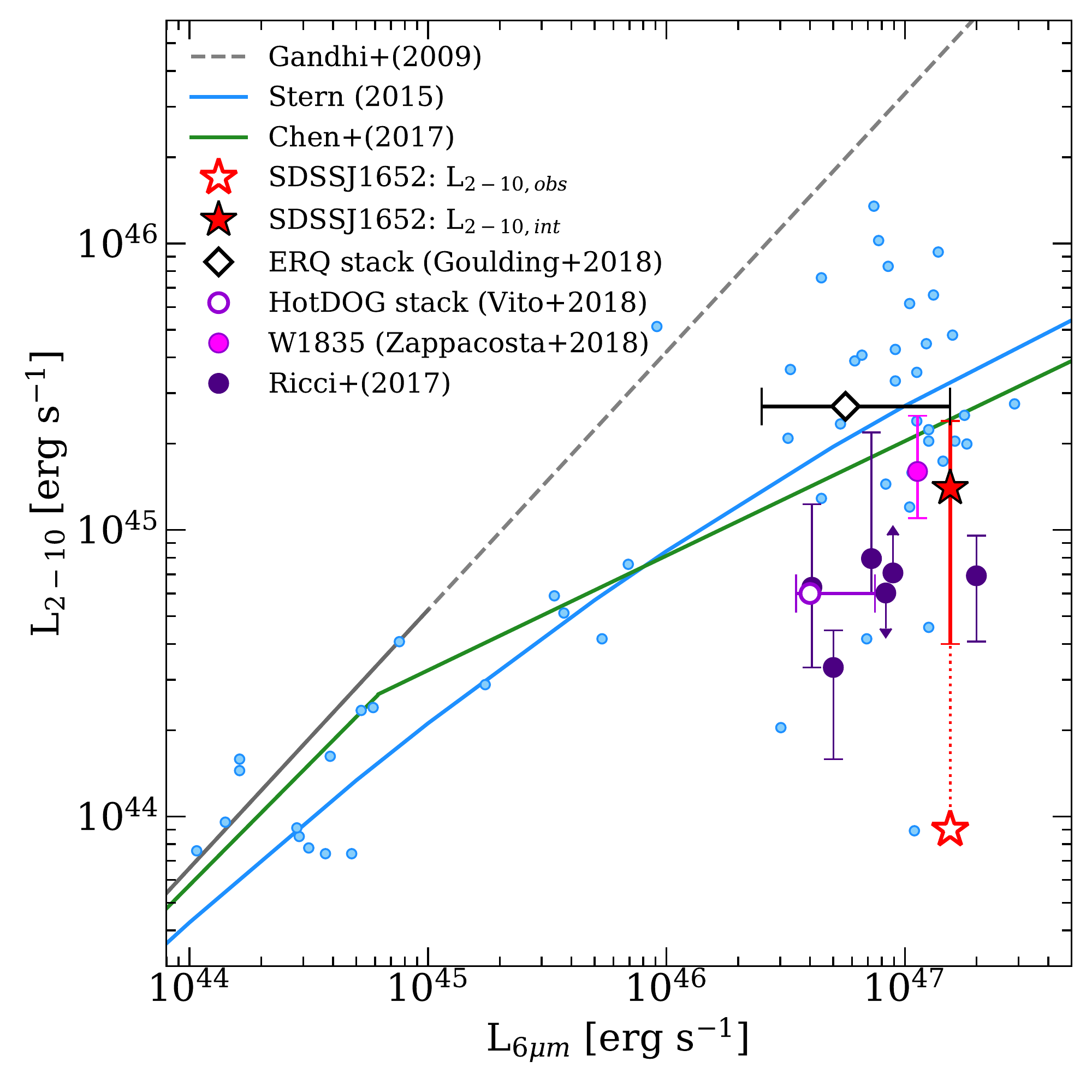}
    	 \end{tabular}
	 \end{center}
	 \caption{The $L_{6\mu\textrm{m}}$-$L_{\textrm{2-10}}$ relationships for selected targets, including \ourERQ\ are shown. All $L_{\textrm{2-10}}$ indicate the absorption corrected, intrinsic X-ray luminosities, unless otherwise noted. Since the original linear relation by \citet{Gandhi2009AA} was derived for $L_{12\mu\textrm{m}}<10^{45}$\ergs, the extrapolated relation at higher luminosities, corrected for $L_{6\mu\textrm{m}}$, is shown as a dashed gray line. Sub-linear relationships derived for higher luminosities by \citet{Stern2015ApJ807129S} and \citet{Chen2017ApJ} are shown as solid blue and green lines; the blue data points correspond to Type I quasars used in \citet{Stern2015ApJ807129S}. The observed (open star) and intrinsic (filled star) $L_{\textrm{2-10}}$ of \ourERQ\ using the \texttt{MYTorus} model is shown with 90\% confidence. The stacked ERQ \citep{Goulding2018} result is shown with the error bars spanning the observed $L_{6\mu\textrm{m}}$ range. Both the stacked ERQ and \ourERQ\ results are consistent with the relationships derived for Type I quasars. We also compare selected HotDOG results \citep{Ricci2017ApJ835}, including stacked analysis showing low $L_{\textrm{2-10}}$ \citep{Vito2018MNRAS474} and a deep observation of W1835+4355 with high $L_{\textrm{2-10}}$ \citep{Zappacosta2018AA}.}
	 \label{fig:LxL6um} 
\end{figure}

X-ray emission and mid-infrared emission are both tracers of black hole accretion activities. X-rays originate from the corona over the inner accretion flow and the mid-infrared emission is re-radiated by the obscuring dust on larger scales. A common analysis is to compare the intrinsic 2-10 keV X-ray luminosity, $L_{\textrm{2-10}}$, against the mid-infrared rest-frame $6\mu\textrm{m}$ luminosity, $L_{6\mu\textrm{m}}$. \citet{Stern2015ApJ807129S} and \citet{Chen2017ApJ} showed an empirically derived sub-linear $L_{6\mu\textrm{m}}$-$L_{\textrm{2-10}}$ relationship for luminous Type I quasars with $ L_{6\mu\textrm{m}} \gtrsim10^{45}$\ergs, in which the measured $L_{\textrm{2-10}}$ at fixed $L_{6\mu\textrm{m}}$ is an order of magnitude below the extrapolations expected from the linear \citet{Gandhi2009AA} relationship, as shown in Figure \ref{fig:LxL6um}. The origin of this sub-linear relationship remain unresolved.

One possibility is that luminous quasars with strong X-rays would overly ionize the circumnuclear gas and suppress the production of radiatively-driven winds \citep{Luo2013ApJ772153L}. The physical connection would then lead to the observed \citet{Stern2015ApJ807129S} and \citet{Chen2017ApJ} relations through the coronal quenching phenomenon \citep{Luo2013ApJ772153L}, in which X-ray emission is suppressed by a massive accretion flow such as a failed disk wind that falls into the accretion disc \citep{Proga2005ApJ630L9P, Leighly2007ApJ663, lusso2010, Jin2012MNRAS}. This would suggest that high accretion efficiency leads to a decrease in radiative efficiency.


Interestingly, there is evidence for other luminous quasars with even weaker observed X-ray luminosities, falling below the \citet{Stern2015ApJ807129S} and \citet{Chen2017ApJ} relations even after accounting for absorption. Examples of quasars with a very weak X-ray emission include a Type I quasar with emission line outflows \citep{Leighly2007ApJ663}, broad absorption line quasars (BAL; \citealt{Luo2013ApJ772153L}), ultraluminous infrared galaxies (ULIRGs; \citealt{Teng2014ApJ,Teng2015ApJ}), and hot dust-obscured galaxies (HotDOGs; \citealt{Ricci2017ApJ835}). Either these quasars are intrinsically X-ray weak or they have a complex inner geometrical structure with multiple absorbers which cannot be identified in a single low-quality spectrum.


Combining the \wise\ photometry data and the X-ray analysis from this paper, we can perform a similar analysis to probe the outflow mechanisms of \ourERQ. In Figure \ref{fig:LxL6um}, we compare the results for \ourERQ\ against the \citet{Stern2015ApJ807129S} and \citet{Chen2017ApJ} relationships and the stacked \chandra\ measurements of ERQs, which includes \ourERQ, by \citet{Goulding2018}. Given the uncertainties of individual sources using the stacking technique, the stacking estimate and our estimate of the intrinsic luminosity of \ourERQ\ are consistent, while deeper observations of \ourERQ\ places a better constraint on the absorption-corrected luminosity. This result supports the \citet{Goulding2018} observation that there is no strong evidence that ERQs are intrinsically X-ray weak: they appear to be consistent with Type I quasars once the intervening absorption is accounted for. These results would suggest that despite the presence of extreme outflows, the heavily obscured \ourERQ\ and other ERQs are not dominated by extreme coronal quenching that would otherwise result in extreme X-ray suppression, hence they are not intrinsically X-ray weak.

It is also possible for the measured $L_{6\mu\textrm{m}}$ to be underestimated due to obscuration or due to anisotopic emission \citep{Zappacosta2018AA}, which would imply \LX\ deficit compared to the standard relationships and strong X-ray suppression. This would require a correction of nearly an order of magnitude to impose a significant departure from the observed $L_{6\mu\textrm{m}}$-$L_{\textrm{2-10}}$ relationship, which is unlikely. \citet{Goulding2018} notes that the observed high luminosities and outflows in ERQs may be due to a combination of color selection and orientation effects. Current observations of \ourERQ\ and other ERQs suggest that they are dissimilar from X-ray weak BAL quasars in \cite{Luo2013ApJ772153L} in terms of their relationship between extreme accretion, coronal quenching, and outflows. Therefore, it is possible that the ERQs have an outflow mechanism that is different from those of BAL quasars.

\begin{figure*}
	 \begin{center}
	 \begin{tabular}{c}
    \includegraphics[trim={0.5cm 3.75cm 2.75cm 3.5cm}, clip, width=0.902\textwidth]{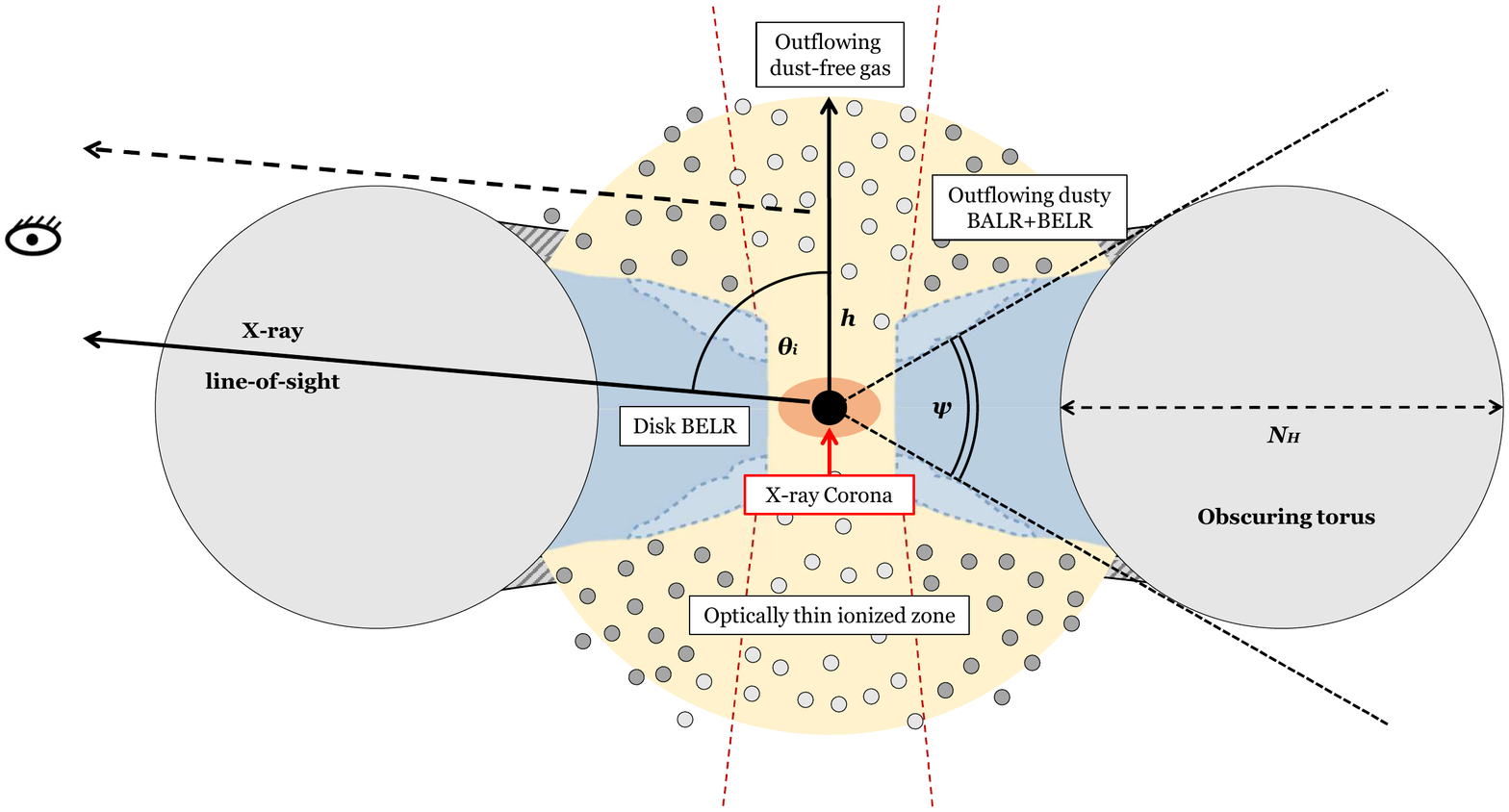}
	 \end{tabular}
	 \end{center}
	 \caption{This diagram (not to scale) shows the preferred model picture for \ourERQ\ adapted from \citet{Alexandroff2018mnras} and combined with our best-fit X-ray \texttt{MYTorus} model parameters described in Table \ref{tab:bestFit}. The reprocessed \texttt{MYTorus} model flux is shown in solid black line, and the scattered flux is in dashed line. This illustrates the cross-sectional view through obscuring torus. The model is inspired by \citet{Veilleux2016ApJ}. The inner accretion disk is surrounded by a geometrically thick BELR accretion disc (in light blue), in which scattering of the dusty equatorial outflow that produces UV-polarization occurs higher in the dusty BALR and BELR. The torus (in gray) geometry is adapted from \citet{MY09MNRAS}, which assumes a half-opening angle of $\psi=60^{\circ}$ (equivalent to a covering factor of $\Delta\Omega/4\pi=0.5$). The best-fitting \texttt{MYTorus} model suggest a highly inclined near edge-on geometry of $\theta_i=85^{\circ}$. Our model describes the X-ray power-law continuum experiencing photoelectric absorption, Compton-scattering, and electron scattering from the warm/hot ionized region surrounding the central engine (in yellow). }
	 \label{fig:cartoonFromalexandroff2018} 
\end{figure*}

\subsection{Obscuration properties}
The obscuration geometry of the best-fit \texttt{MYTorus} model is broadly comprised of two components: the obscuring torus and optically-thin scattering region. The $\theta_i=85^{\circ}$ \texttt{MYTorus} result is consistent with the near-edge-on line-of-sight geometry described by \citet{Alexandroff2018mnras}, which suggest the observed X-ray absorption must occur in same wind as the observed UV scattering \citep{Goulding2018}. In the \citet{Alexandroff2018mnras} model, the polarized UV wind in \ourERQ\ is likely due to dust scattering somewhere in the broad emission line region (BELR) and broad absorption line region (BALR) at scales greater than the obscuration. The inclined viewing angle inferred in the UV is based on the observed shape of the infrared SED. With the inclusion of the X-ray results, our \texttt{MYTorus} model-fits adds the absorbing torus and the warm/hot ionized electron scattering region surrounding the central engine. A cartoon picture of the combined \ourERQ\ model is shown in Figure \ref{fig:cartoonFromalexandroff2018}.

The optically-thin scattered power-law continuum is commonly observed as a rise in the X-ray spectrum towards the low energies. The scattered spectrum is assumed to have the same shape (i.e. $\Gamma$) as the incident power-law spectrum and is assumed to be due to electron scattering. 
This means that the scattering fraction $f_{sc}$ is the ratio of the scattered flux to the incident flux. This extended scattering region needs to be on scales larger than the central engine and the obscuring torus, which in turn must be larger than the derived dust sublimation region. Using the dust sublimation estimate \citep{Barvainis1987ApJ} and the extrapolated UV luminosity, we find that the dust sublimation distance is $\sim1\textrm{ pc}$ \citep{Alexandroff2018mnras}. 

We can describe $f_{sc}$ modeled in \texttt{MYTorus} in terms of the size of the scattering region and the density of the scattering particles \citep{Zakamska2005AJ}. 
We assume scattering is dominated by Thomson electron scattering with scattering cross-section $\sigma_T$ in a region of uniform electron density $n_e$ with differential cross section $d\sigma/d\Omega\sim \sigma_T/4\pi$ through a cone subtending $\Delta\Omega=2\pi$, defined by \texttt{MYTorus} in Figure \ref{fig:cartoonFromalexandroff2018}:
\begin{equation}
    f_{sc} = \frac{d\sigma}{d\Omega} \Delta \Omega \int n_e(r) dr  \sim \frac{1}{2}n_e \sigma_T h 
    \bigg( \frac{\Delta\Omega}{2\pi} \bigg)
        \label{eq:scatfract}
\end{equation}
Interestingly, $f_{sc}\sim3\%$ is similar to the UV scattering estimated from \citet{Alexandroff2018mnras}, which was obtained by extrapolating the IR emission of ERQs using a Type I quasar SED to the observed UV flux. The similar geometry suggested by both X-ray modeling and UV spectropolarimetry, which is dominated by dust scattering, also suggests that the X-ray absorption occurs in the same wind as seen in spectropolarimetric observations. This is consistent with the \citet{Goulding2018} model. Thus, we consider the dust scatter scale height limits $h\sim 1-10\textrm{ pc}$ from \citet{Alexandroff2018mnras} in Equation \ref{eq:scatfract} to calculate the electron scattering density ranging between $n_e\sim7.5\times(10^2-10^3)\textrm{ cm}^{-3}$. 

We also use the scale-height to approximate the line-of-sight mass of the obscuring material. Gas dominates the absorption at X-ray energies \citep{Hickox2018ARAA}, so we consider a hydrogen cloud at fixed metallicity with a constant density $n_H$ distributed over a spherical volume with a scale height $h$, based on the assumption that the scattering zone would be of scales similar to those of obscuration. To match a torus-like distribution, we take the volume $V=\frac{1}{3} h^3 \Delta\Omega$, where $\Delta\Omega=2\pi$ is the solid angle subtended by the torus, based on the \texttt{MYTorus} model in Figure \ref{fig:cartoonFromalexandroff2018}. The resulting gas mass is therefore $M_{obsc}=m_H n_H V$. In terms of the measured column density $N_H=n_H h$, we have
\begin{equation}
    M_{obsc} \sim 
    1.7\times10^{6} M_{\astrosun} 
    \bigg(\frac{N_H}{10^{24}\textrm{ cm}^{-2}}\bigg)  \bigg(\frac{ \Delta\Omega}{2\pi}\bigg) \bigg(\frac{h}{10\textrm{ pc}}\bigg)^2.
\end{equation}
Given the uncertainty in $h$ and in the torus geometry, we calculate the mass range of the line-of-sight obscuration to $M_{obsc}\sim1.7\times(10^4-10^6)M_{\astrosun}$. Although it is more appropriate to properly model the the torus structure, we cannot set any limits on its size from \texttt{MYTorus}. At most, we expect a small correction by a factor of a few.

\subsection{Nature of the \ourERQ}
We calculate the order of magnitude approximations for the central black hole accretion properties. The directly observed $L_{6\mu\textrm{m}}$ of $10^{47.19}$\ergs\ suggests that the bolometric luminosity may be on the order $L_{Bol}\sim10^{47-48}$\ergs, which is in agreement with the \citet{Perrotta2019MNRAS} measurement of $L_{Bol}\sim10^{47.73}$\ergs. This would correspond to a X-ray bolometric correction $\kappa_X\approx380$ for $L_{Bol}=\kappa_X L_{\textrm{2-10},obs}$. This is roughly in agreement with values found for $z=2-4$ hyperluminous Type I quasars \citep{Martocchia2017AA}. 


While there are many scaling relationships for black hole mass $M_{BH}$ measurements (e.g. host stellar mass, $H\beta$ kinematics) there are many caveats that complicate direct comparison to ERQs, as noted by \citet{Zakamska2019MNRAS}. Without reliable X-ray to $M_{BH}$ scaling relations, it is also difficult to independently measure $M_{BH}$ from this study. With this in mind, we consider the $H\beta$ virial mass estimate of the \ourERQ\ supermassive black hole by \citet{Perrotta2019MNRAS}:
$M_{BH}=3\times10^9M_{\astrosun}$, which corresponds to an Eddington luminosity of  $L_{Edd}=4\times10^{47}\textrm{erg s}^{-1}$.
Combined with the \citet{Perrotta2019MNRAS} bolometric luminosity, the Eddington ratio is estimated to be $L_{Bol}/L_{Edd}\sim1.2$. \citet{Zakamska2019MNRAS} explained that a possible argument against high Eddington ratios is an overestimated $L_{Bol}$. However, for obscured populations, underestimated $L_{Bol}$ is more likely especially when the observed $L_{6\mu\textrm{m}}$-$L_{\textrm{2-10}}$ relationship is consistent with \citet{Stern2015ApJ807129S,Chen2017ApJ}. Therefore, \ourERQ\ is consistent with being a near- or super-Eddington, Compton-thick source, exhibiting high $L_{2-10}$ and powerful outflows.

Assuming the material scattering X-rays has an outflowing component, we can also calculate its maximum kinetic energy power using the $n_e$ estimates of the warm scattering region. We assume the same velocities $v\approx3000$ km s$^{-1}$ seen in \citet{Alexandroff2018mnras} launched at scale-height $h=10\textrm{ pc}$:

\begin{equation}
    \begin{split}
        \dot{E}_{sc}  & \approx 4\times10^{45} \textrm{ erg s}^{-1} \\
        &\times \bigg(\frac{n_e}{7.5\times10^3 \textrm{ cm}^{-3}}\bigg) \bigg(\frac{h}{10\textrm{ pc}}\bigg)^2 \bigg(\frac{v}{3000\textrm{ km s}^{-1}}\bigg)^{3}
    \end{split}
\end{equation}
which makes up a small fraction of the bolometric output. Thus, the phase of gas responsible for producing the scattered X-rays is not an energetically important component of the quasar ouput.



\subsection{Comparing with other observations} 

While there are X-ray studies of highly obscured quasars, most have been limited to low-redshift samples. Another class of reddened quasars are the HotDOGs, which have properties similar to those of ERQs in terms of spectral energy distribution, luminosity, and redshift \citep{Eisenhardt2012ApJ755,Assef2015ApJ804}. Despite these similarities, there are some interesting differences stemming from somewhat different selection methods. One such difference is that HotDOGs are
well below the \citet{Stern2015ApJ807129S,Chen2017ApJ} relationships \citep{Ricci2017ApJ835, Vito2018MNRAS474}, while also showing Compton thick obscurations \citep{Piconcelli2015AA}, in contrast to ERQs explored in \citet{Goulding2018} and in this paper. We compare their $L_{6\mu\textrm{m}}$-$L_{\textrm{2-10}}$ properties in Figure \ref{fig:LxL6um}.

A recent \nustar\ observation of a $z=2.298$ HotDOG (W1835+4355) showed Compton-thick obscuration, $N_H\approx 0.9\times10^{24}\textrm{ cm}^{-2}$, and absorption-corrected luminosity of $L_{2-10}=1.6\times10^{45}\textrm{ erg s}^{-1}$ \citep{Zappacosta2018AA} similar to those of \ourERQ\ presented here. Compared to other HotDOGs, this target also exhibited higher X-ray luminosity, nearly consistent with the \citet{Stern2015ApJ807129S,Chen2017ApJ} relations, shown in Figure \ref{fig:LxL6um}. In contrast to \ourERQ, their target was modeled with $\theta_i=70^{\circ}$ and a larger model-dependent $f_{sc}\approx5-15\%$.  This suggest remarkable similarities between ERQs and HotDOGs with the only differences in the obscuration properties: $\theta_i$ and $f_{sc}$. This indicates a variety in HotDOG properties with some that have overlaping $L_{6\mu\textrm{m}}$-$L_{\textrm{2-10}}$ properties to ERQs. 


\section{Conclusion} \label{sec:concl}
We report on deep $\sim$ 130 ks \xmm\ and $\sim$ 218 ks \nustar\ observations of \ourERQ, an obscured, extremely red quasar at $z=2.94$. We characterize the intrinsic X-ray properties and obscuration geometry using \xmm, which is sensitive to the absorbing column, and \nustar, which probes the intrinsically unobscured X-rays. We determine the accretion luminosity and the obscuration properties of \ourERQ\ to provide verification for severely photon-limited studies of the high redshift, obscured quasar population. 

We extracted spectra from the \xmm\ MOS and PN and the \nustar\ FPMA detectors. Unfortunately, we could only place a 3-$\sigma$ upper limit for the \nustar\ FPMB. We fit 0.5 to 10 keV \xmm\ and 6 keV to 79 keV \nustar\ spectra with a physically motivated \texttt{MYTorus} model that models the zeroth order continuum reprocessing due to the intervening cold gas, Compton-scattered continuum, fluorescent line emission, and the optically-thin scattered power-law continuum due to electron scattering in the warm ionized region surrounding the central engine. We find the best-fit model indicates a Compton-thick column density of $N_H=$\ \NHwEH\colden, a near edge-on geometry with the line-of-sight inclination angle of $\theta_i=85^{\circ}$, and a warm electron scattering fraction of $f_{sc}\sim 3\%$. 

From the measured obscuration parameters, we place estimates of the physical properties of the obscuration, assuming scale heights $h\sim1-10$ pc for dust and electron scattering. From $f_{sc}$, we estimate the electron density driving the soft-energy scattering to be between $n_e\sim 7.5\times(10^2-10^3) \textrm{ cm}^{-3}$. From $N_H$ we estimate the obscuration mass to be approximately $M_{obsc}\sim 1.7\times(10^4-10^6) M_{\astrosun}$. These values are highly uncertain due to uncertainties in the torus geometry.

The absorption-corrected, intrinsic X-ray luminosity is \LX$=$\ \LXintFH\ \ergs. This is consistent with the $L_{6\mu\textrm{m}}$-$L_{\textrm{2-10}}$ relationship for luminous Type I quasars, which indicates that \ourERQ\ is not X-ray weak once corrected for the Compton-thick absorption. This trend is consistent with other ERQs, in contrast to other known luminous quasars. The observed $L_{6\mu\textrm{m}}$-$L_{\textrm{2-10}}$ would imply that \ourERQ\ may not be experiencing coronal quenching that would nominally suppress X-ray radiation supposedly associated with high accretion. Moreover, the observed luminosity estimates support the energetics needed to drive the observed galactic scale outflows. These results indicate the ability of near- or super-Eddington, Compton-thick ERQs like \ourERQ\ to drive powerful outflows without requiring intrinsically weak X-rays. 

\section*{Acknowledgements}\label{sec:acknow}
This research was supported by NASA grant NuSTAR GO-4298. This research is based on observations obtained with XMM-Newton, an ESA science mission with instruments and contributions directly funded by ESA Member States and NASA. This research has made use of data obtained with NuSTAR, a project led by Caltech, funded by NASA and managed by NASA/JPL, and has utilized the NUSTARDAS software package, jointly developed by the ASDC (Italy) and Caltech (USA).

We also thank the referee for the many helpful feedback and comments in improving this paper.

\section*{Data Availability} \label{sec:dataAvail}
This research has made use of data and/or software provided by the High Energy Astrophysics Science Archive Research Center (HEASARC), which is a service of the Astrophysics Science Division at NASA/GSFC. The data underlying this article are available in the HEASARC Archives at  \texttt{https://heasarc.gsfc.nasa.gov/} and the XMM-Newton Science Archives at \texttt{http://nxsa.esac.esa.int/}.



\bibliographystyle{mnras}
\bibliography{xray} 

\begin{thebibliography}{}
\makeatletter
\relax
\def\mn@urlcharsother{\let\do\@makeother \do\$\do\&\do\#\do\^\do\_\do\%\do\~}
\def\mn@doi{\begingroup\mn@urlcharsother \@ifnextchar [ {\mn@doi@}
  {\mn@doi@[]}}
\def\mn@doi@[#1]#2{\def\@tempa{#1}\ifx\@tempa\@empty \href
  {http://dx.doi.org/#2} {doi:#2}\else \href {http://dx.doi.org/#2} {#1}\fi
  \endgroup}
\def\mn@eprint#1#2{\mn@eprint@#1:#2::\@nil}
\def\mn@eprint@arXiv#1{\href {http://arxiv.org/abs/#1} {{\tt arXiv:#1}}}
\def\mn@eprint@dblp#1{\href {http://dblp.uni-trier.de/rec/bibtex/#1.xml}
  {dblp:#1}}
\def\mn@eprint@#1:#2:#3:#4\@nil{\def\@tempa {#1}\def\@tempb {#2}\def\@tempc
  {#3}\ifx \@tempc \@empty \let \@tempc \@tempb \let \@tempb \@tempa \fi \ifx
  \@tempb \@empty \def\@tempb {arXiv}\fi \@ifundefined
  {mn@eprint@\@tempb}{\@tempb:\@tempc}{\expandafter \expandafter \csname
  mn@eprint@\@tempb\endcsname \expandafter{\@tempc}}}

\bibitem[\protect\citeauthoryear{{Ahn} et~al.,}{{Ahn}
  et~al.}{2012}]{sdss2012DR9}
{Ahn} C.~P.,  et~al., 2012, \mn@doi [\apjs] {10.1088/0067-0049/203/2/21}, \href
  {https://ui.adsabs.harvard.edu/abs/2012ApJS..203...21A} {203, 21}

\bibitem[\protect\citeauthoryear{{Alexander}, {Brandt}, {Hornschemeier},
  {Garmire}, {Schneider}, {Bauer}  \& {Griffiths}}{{Alexander}
  et~al.}{2001}]{Alexander2001AJ}
{Alexander} D.~M.,  {Brandt} W.~N.,  {Hornschemeier} A.~E.,  {Garmire} G.~P.,
  {Schneider} D.~P.,  {Bauer} F.~E.,   {Griffiths} R.~E.,  2001, \mn@doi [\aj]
  {10.1086/323540}, \href
  {https://ui.adsabs.harvard.edu/abs/2001AJ....122.2156A} {122, 2156}

\bibitem[\protect\citeauthoryear{{Alexandroff} et~al.,}{{Alexandroff}
  et~al.}{2018}]{Alexandroff2018mnras}
{Alexandroff} R.~M.,  et~al., 2018, \mn@doi [\mnras] {10.1093/mnras/sty1685},
  \href {https://ui.adsabs.harvard.edu/abs/2018MNRAS.479.4936A} {479, 4936}

\bibitem[\protect\citeauthoryear{{Antonucci}}{{Antonucci}}{1993}]{Antonucci1993ARAA}
{Antonucci} R.,  1993, \mn@doi [\araa] {10.1146/annurev.aa.31.090193.002353},
  \href {https://ui.adsabs.harvard.edu/abs/1993ARA&A..31..473A} {31, 473}

\bibitem[\protect\citeauthoryear{{Arnaud}}{{Arnaud}}{1996}]{Arnaud1996ASPC}
{Arnaud} K.~A.,  1996, in {Jacoby} G.~H.,  {Barnes} J.,  eds,  Astronomical
  Society of the Pacific Conference Series Vol. 101, Astronomical Data Analysis
  Software and Systems V. p.~17

\bibitem[\protect\citeauthoryear{{Assef} et~al.,}{{Assef}
  et~al.}{2015}]{Assef2015ApJ804}
{Assef} R.~J.,  et~al., 2015, \mn@doi [\apj] {10.1088/0004-637X/804/1/27},
  \href {https://ui.adsabs.harvard.edu/abs/2015ApJ...804...27A} {804, 27}

\bibitem[\protect\citeauthoryear{{Balokovi{\'c}} et~al.,}{{Balokovi{\'c}}
  et~al.}{2018}]{Balokovic2018}
{Balokovi{\'c}} M.,  et~al., 2018, \mn@doi [\apj] {10.3847/1538-4357/aaa7eb},
  \href {https://ui.adsabs.harvard.edu/abs/2018ApJ...854...42B} {854, 42}

\bibitem[\protect\citeauthoryear{{Barnes} \& {Hernquist}}{{Barnes} \&
  {Hernquist}}{1992}]{BarnesHernquist1992}
{Barnes} J.~E.,  {Hernquist} L.,  1992, \mn@doi [\araa]
  {10.1146/annurev.aa.30.090192.003421}, \href
  {https://ui.adsabs.harvard.edu/abs/1992ARA&A..30..705B} {30, 705}

\bibitem[\protect\citeauthoryear{{Barvainis}}{{Barvainis}}{1987}]{Barvainis1987ApJ}
{Barvainis} R.,  1987, \mn@doi [\apj] {10.1086/165571}, \href
  {https://ui.adsabs.harvard.edu/abs/1987ApJ...320..537B} {320, 537}

\bibitem[\protect\citeauthoryear{{Boyle} \& {Terlevich}}{{Boyle} \&
  {Terlevich}}{1998}]{Boyle&Terlevich1998MNRAS}
{Boyle} B.~J.,  {Terlevich} R.~J.,  1998, \mn@doi [\mnras]
  {10.1046/j.1365-8711.1998.01264.x}, \href
  {https://ui.adsabs.harvard.edu/abs/1998MNRAS.293L..49B} {293, L49}

\bibitem[\protect\citeauthoryear{{Cash}}{{Cash}}{1979}]{Cash1979ApJ}
{Cash} W.,  1979, \mn@doi [\apj] {10.1086/156922}, \href
  {https://ui.adsabs.harvard.edu/abs/1979ApJ...228..939C} {228, 939}

\bibitem[\protect\citeauthoryear{{Chen} et~al.,}{{Chen}
  et~al.}{2017}]{Chen2017ApJ}
{Chen} C.-T.~J.,  et~al., 2017, \mn@doi [\apj] {10.3847/1538-4357/837/2/145},
  \href {https://ui.adsabs.harvard.edu/abs/2017ApJ...837..145C} {837, 145}

\bibitem[\protect\citeauthoryear{{Croton} et~al.,}{{Croton}
  et~al.}{2006}]{Croton2006MNRAS}
{Croton} D.~J.,  et~al., 2006, \mn@doi [\mnras]
  {10.1111/j.1365-2966.2005.09675.x}, \href
  {https://ui.adsabs.harvard.edu/abs/2006MNRAS.365...11C} {365, 11}

\bibitem[\protect\citeauthoryear{{Eisenhardt} et~al.,}{{Eisenhardt}
  et~al.}{2012}]{Eisenhardt2012ApJ755}
{Eisenhardt} P. R.~M.,  et~al., 2012, \mn@doi [\apj]
  {10.1088/0004-637X/755/2/173}, \href
  {https://ui.adsabs.harvard.edu/abs/2012ApJ...755..173E} {755, 173}

\bibitem[\protect\citeauthoryear{{Eisenstein} et~al.,}{{Eisenstein}
  et~al.}{2011}]{sdss2011AJ}
{Eisenstein} D.~J.,  et~al., 2011, \mn@doi [\aj] {10.1088/0004-6256/142/3/72},
  \href {https://ui.adsabs.harvard.edu/abs/2011AJ....142...72E} {142, 72}

\bibitem[\protect\citeauthoryear{{Fabian}}{{Fabian}}{2012}]{Fabian2012ARAA}
{Fabian} A.~C.,  2012, \mn@doi [\araa] {10.1146/annurev-astro-081811-125521},
  \href {https://ui.adsabs.harvard.edu/abs/2012ARA&A..50..455F} {50, 455}

\bibitem[\protect\citeauthoryear{{Fischer} et~al.,}{{Fischer}
  et~al.}{2018}]{Fischer2018ApJ856}
{Fischer} T.~C.,  et~al., 2018, \mn@doi [\apj] {10.3847/1538-4357/aab03e},
  \href {https://ui.adsabs.harvard.edu/abs/2018ApJ...856..102F} {856, 102}

\bibitem[\protect\citeauthoryear{{Gabriel} et~al.,}{{Gabriel}
  et~al.}{2004}]{Gabriel2004xmmsas}
{Gabriel} C.,  et~al., 2004, in {Ochsenbein} F.,  {Allen} M.~G.,   {Egret} D.,
  eds,  Astronomical Society of the Pacific Conference Series Vol. 314,
  Astronomical Data Analysis Software and Systems (ADASS) XIII. p.~759

\bibitem[\protect\citeauthoryear{{Gaia Collaboration}}{{Gaia
  Collaboration}}{2018}]{Gaia2018yCat}
{Gaia Collaboration} 2018, VizieR Online Data Catalog, \href
  {https://ui.adsabs.harvard.edu/abs/2018yCat.1345....0G} {p. I/345}

\bibitem[\protect\citeauthoryear{{Gandhi}, {Horst}, {Smette}, {H{\"o}nig},
  {Comastri}, {Gilli}, {Vignali}  \& {Duschl}}{{Gandhi}
  et~al.}{2009}]{Gandhi2009AA}
{Gandhi} P.,  {Horst} H.,  {Smette} A.,  {H{\"o}nig} S.,  {Comastri} A.,
  {Gilli} R.,  {Vignali} C.,   {Duschl} W.,  2009, \mn@doi [\aap]
  {10.1051/0004-6361/200811368}, \href
  {https://ui.adsabs.harvard.edu/abs/2009A&A...502..457G} {502, 457}

\bibitem[\protect\citeauthoryear{Goulding et~al.,}{Goulding
  et~al.}{2018}]{Goulding2018}
Goulding A.~D.,  et~al., 2018, \mn@doi [\apj] {10.3847/1538-4357/aab040}, 856,
  4

\bibitem[\protect\citeauthoryear{{Greene}, {Zakamska}  \& {Smith}}{{Greene}
  et~al.}{2012}]{Greene2012ApJ746}
{Greene} J.~E.,  {Zakamska} N.~L.,   {Smith} P.~S.,  2012, \mn@doi [\apj]
  {10.1088/0004-637X/746/1/86}, \href
  {https://ui.adsabs.harvard.edu/abs/2012ApJ...746...86G} {746, 86}

\bibitem[\protect\citeauthoryear{{HI4PI Collaboration} et~al.,}{{HI4PI
  Collaboration} et~al.}{2016}]{HI4PI2016AA}
{HI4PI Collaboration} et~al., 2016, \mn@doi [\aap]
  {10.1051/0004-6361/201629178}, \href
  {https://ui.adsabs.harvard.edu/abs/2016A&A...594A.116H} {594, A116}

\bibitem[\protect\citeauthoryear{{Hamann} et~al.,}{{Hamann}
  et~al.}{2017}]{Hamann2017MNRAS}
{Hamann} F.,  et~al., 2017, \mn@doi [\mnras] {10.1093/mnras/stw2387}, \href
  {https://ui.adsabs.harvard.edu/abs/2017MNRAS.464.3431H} {464, 3431}

\bibitem[\protect\citeauthoryear{{Harrison} et~al.,}{{Harrison}
  et~al.}{2013}]{Harrison2013ApJ770}
{Harrison} F.~A.,  et~al., 2013, \mn@doi [\apj] {10.1088/0004-637X/770/2/103},
  \href {https://ui.adsabs.harvard.edu/abs/2013ApJ...770..103H} {770, 103}

\bibitem[\protect\citeauthoryear{{Harrison}, {Costa}, {Tadhunter},
  {Fl{\"u}tsch}, {Kakkad}, {Perna}  \& {Vietri}}{{Harrison}
  et~al.}{2018}]{Harrison2018NatAs}
{Harrison} C.~M.,  {Costa} T.,  {Tadhunter} C.~N.,  {Fl{\"u}tsch} A.,  {Kakkad}
  D.,  {Perna} M.,   {Vietri} G.,  2018, \mn@doi [Nature Astronomy]
  {10.1038/s41550-018-0403-6}, \href
  {https://ui.adsabs.harvard.edu/abs/2018NatAs...2..198H} {2, 198}

\bibitem[\protect\citeauthoryear{{Hickox} \& {Alexander}}{{Hickox} \&
  {Alexander}}{2018}]{Hickox2018ARAA}
{Hickox} R.~C.,  {Alexander} D.~M.,  2018, \mn@doi [\araa]
  {10.1146/annurev-astro-081817-051803}, \href
  {https://ui.adsabs.harvard.edu/abs/2018ARA&A..56..625H} {56, 625}

\bibitem[\protect\citeauthoryear{{Hopkins} \& {Hernquist}}{{Hopkins} \&
  {Hernquist}}{2009}]{Hopkins2009ApJ694}
{Hopkins} P.~F.,  {Hernquist} L.,  2009, \mn@doi [\apj]
  {10.1088/0004-637X/694/1/599}, \href
  {https://ui.adsabs.harvard.edu/abs/2009ApJ...694..599H} {694, 599}

\bibitem[\protect\citeauthoryear{{Hopkins}, {Hernquist}, {Cox}, {Di Matteo},
  {Robertson}  \& {Springel}}{{Hopkins} et~al.}{2006}]{Hopkins2006ApJS}
{Hopkins} P.~F.,  {Hernquist} L.,  {Cox} T.~J.,  {Di Matteo} T.,  {Robertson}
  B.,   {Springel} V.,  2006, \mn@doi [\apjs] {10.1086/499298}, \href
  {https://ui.adsabs.harvard.edu/abs/2006ApJS..163....1H} {163, 1}

\bibitem[\protect\citeauthoryear{{Jansen} et~al.,}{{Jansen}
  et~al.}{2001}]{Jansen2001AA}
{Jansen} F.,  et~al., 2001, \mn@doi [\aap] {10.1051/0004-6361:20000036}, \href
  {https://ui.adsabs.harvard.edu/abs/2001A&A...365L...1J} {365, L1}

\bibitem[\protect\citeauthoryear{{Jin}, {Ward}  \& {Done}}{{Jin}
  et~al.}{2012}]{Jin2012MNRAS}
{Jin} C.,  {Ward} M.,   {Done} C.,  2012, \mn@doi [\mnras]
  {10.1111/j.1365-2966.2012.21272.x}, \href
  {https://ui.adsabs.harvard.edu/abs/2012MNRAS.425..907J} {425, 907}

\bibitem[\protect\citeauthoryear{{Kormendy} \& {Ho}}{{Kormendy} \&
  {Ho}}{2013}]{KormendyHo2013ARAA}
{Kormendy} J.,  {Ho} L.~C.,  2013, \mn@doi [\araa]
  {10.1146/annurev-astro-082708-101811}, \href
  {https://ui.adsabs.harvard.edu/abs/2013ARA&A..51..511K} {51, 511}

\bibitem[\protect\citeauthoryear{{LaMassa}, {Yaqoob}, {Boorman}, {Tzanavaris},
  {Levenson}, {Gandhi}, {Ptak}  \& {Heckman}}{{LaMassa}
  et~al.}{2019}]{LaMassa2019ApJ}
{LaMassa} S.~M.,  {Yaqoob} T.,  {Boorman} P.~G.,  {Tzanavaris} P.,  {Levenson}
  N.~A.,  {Gandhi} P.,  {Ptak} A.~F.,   {Heckman} T.~M.,  2019, \mn@doi [\apj]
  {10.3847/1538-4357/ab552c}, \href
  {https://ui.adsabs.harvard.edu/abs/2019ApJ...887..173L} {887, 173}

\bibitem[\protect\citeauthoryear{{Lambrides}, {Chiaberge}, {Heckman}, {Gilli},
  {Vito}  \& {Norman}}{{Lambrides} et~al.}{2020}]{Lambrides2020ApJ}
{Lambrides} E.~L.,  {Chiaberge} M.,  {Heckman} T.,  {Gilli} R.,  {Vito} F.,
  {Norman} C.,  2020, \mn@doi [\apj] {10.3847/1538-4357/ab919c}, \href
  {https://ui.adsabs.harvard.edu/abs/2020ApJ...897..160L} {897, 160}

\bibitem[\protect\citeauthoryear{{Lawrence} \& {Elvis}}{{Lawrence} \&
  {Elvis}}{2010}]{LawrenceElvis2010ApJ}
{Lawrence} A.,  {Elvis} M.,  2010, \mn@doi [\apj]
  {10.1088/0004-637X/714/1/561}, \href
  {https://ui.adsabs.harvard.edu/abs/2010ApJ...714..561L} {714, 561}

\bibitem[\protect\citeauthoryear{{Leighly}, {Halpern}, {Jenkins}, {Grupe},
  {Choi}  \& {Prescott}}{{Leighly} et~al.}{2007}]{Leighly2007ApJ663}
{Leighly} K.~M.,  {Halpern} J.~P.,  {Jenkins} E.~B.,  {Grupe} D.,  {Choi} J.,
  {Prescott} K.~B.,  2007, \mn@doi [\apj] {10.1086/518017}, \href
  {https://ui.adsabs.harvard.edu/abs/2007ApJ...663..103L} {663, 103}

\bibitem[\protect\citeauthoryear{{Liu}, {Zakamska}, {Greene}, {Nesvadba}  \&
  {Liu}}{{Liu} et~al.}{2013a}]{Liu2013MNRASa}
{Liu} G.,  {Zakamska} N.~L.,  {Greene} J.~E.,  {Nesvadba} N. P.~H.,   {Liu} X.,
   2013a, \mn@doi [\mnras] {10.1093/mnras/stt051}, \href
  {https://ui.adsabs.harvard.edu/abs/2013MNRAS.430.2327L} {430, 2327}

\bibitem[\protect\citeauthoryear{{Liu}, {Zakamska}, {Greene}, {Nesvadba}  \&
  {Liu}}{{Liu} et~al.}{2013b}]{Liu2013MNRASb}
{Liu} G.,  {Zakamska} N.~L.,  {Greene} J.~E.,  {Nesvadba} N. P.~H.,   {Liu} X.,
   2013b, \mn@doi [\mnras] {10.1093/mnras/stt1755}, \href
  {https://ui.adsabs.harvard.edu/abs/2013MNRAS.436.2576L} {436, 2576}

\bibitem[\protect\citeauthoryear{{Luo} et~al.,}{{Luo}
  et~al.}{2013}]{Luo2013ApJ772153L}
{Luo} B.,  et~al., 2013, \mn@doi [\apj] {10.1088/0004-637X/772/2/153}, \href
  {https://ui.adsabs.harvard.edu/abs/2013ApJ...772..153L} {772, 153}

\bibitem[\protect\citeauthoryear{{Lusso} et~al.,}{{Lusso}
  et~al.}{2010}]{lusso2010}
{Lusso} E.,  et~al., 2010, \mn@doi [\aap] {10.1051/0004-6361/200913298}, \href
  {https://ui.adsabs.harvard.edu/abs/2010A&A...512A..34L} {512, A34}

\bibitem[\protect\citeauthoryear{{Lusso} et~al.,}{{Lusso}
  et~al.}{2013}]{Lusso2013ApJ}
{Lusso} E.,  et~al., 2013, \mn@doi [\apj] {10.1088/0004-637X/777/2/86}, \href
  {https://ui.adsabs.harvard.edu/abs/2013ApJ...777...86L} {777, 86}

\bibitem[\protect\citeauthoryear{{Martocchia} et~al.,}{{Martocchia}
  et~al.}{2017}]{Martocchia2017AA}
{Martocchia} S.,  et~al., 2017, \mn@doi [\aap] {10.1051/0004-6361/201731314},
  \href {https://ui.adsabs.harvard.edu/abs/2017A&A...608A..51M} {608, A51}

\bibitem[\protect\citeauthoryear{{Murphy} \& {Yaqoob}}{{Murphy} \&
  {Yaqoob}}{2009}]{MY09MNRAS}
{Murphy} K.~D.,  {Yaqoob} T.,  2009, \mn@doi [\mnras]
  {10.1111/j.1365-2966.2009.15025.x}, \href
  {https://ui.adsabs.harvard.edu/abs/2009MNRAS.397.1549M} {397, 1549}

\bibitem[\protect\citeauthoryear{{Murray}, {Chiang}, {Grossman}  \&
  {Voit}}{{Murray} et~al.}{1995}]{Murray1995ApJ451}
{Murray} N.,  {Chiang} J.,  {Grossman} S.~A.,   {Voit} G.~M.,  1995, \mn@doi
  [\apj] {10.1086/176238}, \href
  {https://ui.adsabs.harvard.edu/abs/1995ApJ...451..498M} {451, 498}

\bibitem[\protect\citeauthoryear{{Nandra} \& {Pounds}}{{Nandra} \&
  {Pounds}}{1994}]{NandraPounds1994MNRAS}
{Nandra} K.,  {Pounds} K.~A.,  1994, \mn@doi [\mnras]
  {10.1093/mnras/268.2.405}, \href
  {https://ui.adsabs.harvard.edu/abs/1994MNRAS.268..405N} {268, 405}

\bibitem[\protect\citeauthoryear{{Page}, {Reeves}, {O'Brien}  \&
  {Turner}}{{Page} et~al.}{2005}]{Page2005MNRAS}
{Page} K.~L.,  {Reeves} J.~N.,  {O'Brien} P.~T.,   {Turner} M.~J.~L.,  2005,
  \mn@doi [\mnras] {10.1111/j.1365-2966.2005.09550.x}, \href
  {https://ui.adsabs.harvard.edu/abs/2005MNRAS.364..195P} {364, 195}

\bibitem[\protect\citeauthoryear{{Perrotta}, {Hamann}, {Zakamska}, {Alexand
  roff}, {Rupke}  \& {Wylezalek}}{{Perrotta} et~al.}{2019}]{Perrotta2019MNRAS}
{Perrotta} S.,  {Hamann} F.,  {Zakamska} N.~L.,  {Alexand roff} R.~M.,  {Rupke}
  D.,   {Wylezalek} D.,  2019, \mn@doi [\mnras] {10.1093/mnras/stz1993}, \href
  {https://ui.adsabs.harvard.edu/abs/2019MNRAS.488.4126P} {488, 4126}

\bibitem[\protect\citeauthoryear{{Piconcelli}, {Jimenez-Bail{\'o}n},
  {Guainazzi}, {Schartel}, {Rodr{\'\i}guez-Pascual}  \&
  {Santos-Lle{\'o}}}{{Piconcelli} et~al.}{2005}]{Piconcelli2005AA}
{Piconcelli} E.,  {Jimenez-Bail{\'o}n} E.,  {Guainazzi} M.,  {Schartel} N.,
  {Rodr{\'\i}guez-Pascual} P.~M.,   {Santos-Lle{\'o}} M.,  2005, \mn@doi [\aap]
  {10.1051/0004-6361:20041621}, \href
  {https://ui.adsabs.harvard.edu/abs/2005A&A...432...15P} {432, 15}

\bibitem[\protect\citeauthoryear{{Piconcelli} et~al.,}{{Piconcelli}
  et~al.}{2015}]{Piconcelli2015AA}
{Piconcelli} E.,  et~al., 2015, \mn@doi [\aap] {10.1051/0004-6361/201425324},
  \href {https://ui.adsabs.harvard.edu/abs/2015A&A...574L...9P} {574, L9}

\bibitem[\protect\citeauthoryear{{Proga}}{{Proga}}{2005}]{Proga2005ApJ630L9P}
{Proga} D.,  2005, \mn@doi [\apjl] {10.1086/462417}, \href
  {https://ui.adsabs.harvard.edu/abs/2005ApJ...630L...9P} {630, L9}

\bibitem[\protect\citeauthoryear{{Proga}, {Stone}  \& {Kallman}}{{Proga}
  et~al.}{2000}]{Proga2000ApJ543}
{Proga} D.,  {Stone} J.~M.,   {Kallman} T.~R.,  2000, \mn@doi [\apj]
  {10.1086/317154}, \href
  {https://ui.adsabs.harvard.edu/abs/2000ApJ...543..686P} {543, 686}

\bibitem[\protect\citeauthoryear{{Reeves} \& {Turner}}{{Reeves} \&
  {Turner}}{2000}]{Reeves2000MNRAS}
{Reeves} J.~N.,  {Turner} M.~J.~L.,  2000, \mn@doi [\mnras]
  {10.1046/j.1365-8711.2000.03510.x}, \href
  {https://ui.adsabs.harvard.edu/abs/2000MNRAS.316..234R} {316, 234}

\bibitem[\protect\citeauthoryear{{Ricci} et~al.,}{{Ricci}
  et~al.}{2017}]{Ricci2017ApJ835}
{Ricci} C.,  et~al., 2017, \mn@doi [\apj] {10.3847/1538-4357/835/1/105}, \href
  {https://ui.adsabs.harvard.edu/abs/2017ApJ...835..105R} {835, 105}

\bibitem[\protect\citeauthoryear{{Ross} et~al.,}{{Ross}
  et~al.}{2015}]{Ross2015MNRAS}
{Ross} N.~P.,  et~al., 2015, \mn@doi [\mnras] {10.1093/mnras/stv1710}, \href
  {https://ui.adsabs.harvard.edu/abs/2015MNRAS.453.3932R} {453, 3932}

\bibitem[\protect\citeauthoryear{{Sanders} \& {Mirabel}}{{Sanders} \&
  {Mirabel}}{1996}]{SandersMirabel1996}
{Sanders} D.~B.,  {Mirabel} I.~F.,  1996, \mn@doi [\araa]
  {10.1146/annurev.astro.34.1.749}, \href
  {https://ui.adsabs.harvard.edu/abs/1996ARA&A..34..749S} {34, 749}

\bibitem[\protect\citeauthoryear{{Sanders}, {Soifer}, {Elias}, {Madore},
  {Matthews}, {Neugebauer}  \& {Scoville}}{{Sanders}
  et~al.}{1988}]{Sanders1988ApJ}
{Sanders} D.~B.,  {Soifer} B.~T.,  {Elias} J.~H.,  {Madore} B.~F.,  {Matthews}
  K.,  {Neugebauer} G.,   {Scoville} N.~Z.,  1988, \mn@doi [\apj]
  {10.1086/165983}, \href
  {https://ui.adsabs.harvard.edu/abs/1988ApJ...325...74S} {325, 74}

\bibitem[\protect\citeauthoryear{{Silk} \& {Rees}}{{Silk} \&
  {Rees}}{1998}]{SilkRees1998AA}
{Silk} J.,  {Rees} M.~J.,  1998, \aap, \href
  {https://ui.adsabs.harvard.edu/abs/1998A&A...331L...1S} {331, L1}

\bibitem[\protect\citeauthoryear{{Stern}}{{Stern}}{2015}]{Stern2015ApJ807129S}
{Stern} D.,  2015, \mn@doi [\apj] {10.1088/0004-637X/807/2/129}, \href
  {https://ui.adsabs.harvard.edu/abs/2015ApJ...807..129S} {807, 129}

\bibitem[\protect\citeauthoryear{{Str{\"u}der} et~al.,}{{Str{\"u}der}
  et~al.}{2001}]{struder2001AA}
{Str{\"u}der} L.,  et~al., 2001, \mn@doi [\aap] {10.1051/0004-6361:20000066},
  \href {https://ui.adsabs.harvard.edu/abs/2001A&A...365L..18S} {365, L18}

\bibitem[\protect\citeauthoryear{{Teng} et~al.,}{{Teng}
  et~al.}{2014}]{Teng2014ApJ}
{Teng} S.~H.,  et~al., 2014, \mn@doi [\apj] {10.1088/0004-637X/785/1/19}, \href
  {https://ui.adsabs.harvard.edu/abs/2014ApJ...785...19T} {785, 19}

\bibitem[\protect\citeauthoryear{{Teng} et~al.,}{{Teng}
  et~al.}{2015}]{Teng2015ApJ}
{Teng} S.~H.,  et~al., 2015, \mn@doi [\apj] {10.1088/0004-637X/814/1/56}, \href
  {https://ui.adsabs.harvard.edu/abs/2015ApJ...814...56T} {814, 56}

\bibitem[\protect\citeauthoryear{{Turner} et~al.,}{{Turner}
  et~al.}{2001}]{Turner2001AA}
{Turner} M.~J.~L.,  et~al., 2001, \mn@doi [\aap] {10.1051/0004-6361:20000087},
  \href {https://ui.adsabs.harvard.edu/abs/2001A&A...365L..27T} {365, L27}

\bibitem[\protect\citeauthoryear{{Ueda}, {Akiyama}, {Ohta}  \& {Miyaji}}{{Ueda}
  et~al.}{2003}]{Ueda2003ApJ}
{Ueda} Y.,  {Akiyama} M.,  {Ohta} K.,   {Miyaji} T.,  2003, \mn@doi [\apj]
  {10.1086/378940}, \href
  {https://ui.adsabs.harvard.edu/abs/2003ApJ...598..886U} {598, 886}

\bibitem[\protect\citeauthoryear{{Vayner} et~al.,}{{Vayner}
  et~al.}{2021}]{vayner20}
{Vayner} A.,  et~al., 2021, \mnras, submitted

\bibitem[\protect\citeauthoryear{{Veilleux}, {Mel{\'e}ndez}, {Tripp}, {Hamann}
  \& {Rupke}}{{Veilleux} et~al.}{2016}]{Veilleux2016ApJ}
{Veilleux} S.,  {Mel{\'e}ndez} M.,  {Tripp} T.~M.,  {Hamann} F.,   {Rupke}
  D.~S.~N.,  2016, \mn@doi [\apj] {10.3847/0004-637X/825/1/42}, \href
  {https://ui.adsabs.harvard.edu/abs/2016ApJ...825...42V} {825, 42}

\bibitem[\protect\citeauthoryear{{Vito} et~al.,}{{Vito}
  et~al.}{2018}]{Vito2018MNRAS474}
{Vito} F.,  et~al., 2018, \mn@doi [\mnras] {10.1093/mnras/stx3120}, \href
  {https://ui.adsabs.harvard.edu/abs/2018MNRAS.474.4528V} {474, 4528}

\bibitem[\protect\citeauthoryear{{Wright} et~al.,}{{Wright}
  et~al.}{2010}]{wright2010}
{Wright} E.~L.,  et~al., 2010, \mn@doi [\aj] {10.1088/0004-6256/140/6/1868},
  \href {https://ui.adsabs.harvard.edu/abs/2010AJ....140.1868W} {140, 1868}

\bibitem[\protect\citeauthoryear{{Wylezalek}, {Barrera-Ballesteros},
  {Luetzgendorf}, {Nesvadba}, {Rupke}, {Sun}, {Veilleux}  \&
  {Zakamska}}{{Wylezalek} et~al.}{2017}]{q3d2017jwst}
{Wylezalek} D.,  {Barrera-Ballesteros} J.~K.,  {Luetzgendorf} N.,  {Nesvadba}
  N.,  {Rupke} D.,  {Sun} A.-L.,  {Veilleux} S.,   {Zakamska} N.~L.,  2017,
  {Q-3D: Imaging Spectroscopy of Quasar Hosts with JWST Analyzed with a
  Powerful New PSF Decomposition and Spectral Analysis Package}, JWST Proposal
  ID 1335. Cycle 0 Early Release Science

\bibitem[\protect\citeauthoryear{{Yaqoob}}{{Yaqoob}}{2012}]{Yaqoob2012MNRAS}
{Yaqoob} T.,  2012, \mn@doi [\mnras] {10.1111/j.1365-2966.2012.21129.x}, \href
  {https://ui.adsabs.harvard.edu/abs/2012MNRAS.423.3360Y} {423, 3360}

\bibitem[\protect\citeauthoryear{{Zakamska} et~al.,}{{Zakamska}
  et~al.}{2005}]{Zakamska2005AJ}
{Zakamska} N.~L.,  et~al., 2005, \mn@doi [\aj] {10.1086/427543}, \href
  {https://ui.adsabs.harvard.edu/abs/2005AJ....129.1212Z} {129, 1212}

\bibitem[\protect\citeauthoryear{{Zakamska} et~al.,}{{Zakamska}
  et~al.}{2016}]{Zakamska2016MNRAS4593144Z}
{Zakamska} N.~L.,  et~al., 2016, \mn@doi [\mnras] {10.1093/mnras/stw718}, \href
  {https://ui.adsabs.harvard.edu/abs/2016MNRAS.459.3144Z} {459, 3144}

\bibitem[\protect\citeauthoryear{{Zakamska} et~al.,}{{Zakamska}
  et~al.}{2019}]{Zakamska2019MNRAS}
{Zakamska} N.~L.,  et~al., 2019, \mn@doi [\mnras] {10.1093/mnras/stz2071},
  \href {https://ui.adsabs.harvard.edu/abs/2019MNRAS.489..497Z} {489, 497}

\bibitem[\protect\citeauthoryear{{Zappacosta} et~al.,}{{Zappacosta}
  et~al.}{2018}]{Zappacosta2018AA}
{Zappacosta} L.,  et~al., 2018, \mn@doi [\aap] {10.1051/0004-6361/201732557},
  \href {https://ui.adsabs.harvard.edu/abs/2018A&A...618A..28Z} {618, A28}

\makeatother
\end{thebibliography}








\bsp	
\label{lastpage}
\end{document}